\documentclass[twocolumn]{emulateapj}
\usepackage{graphicx}
\usepackage{amssymb}
\usepackage{natbib}
\usepackage{hyperref}
\hypersetup{
  colorlinks=true,
  urlcolor=blue,
  citecolor=blue,
  linkcolor=red
}
\def \eg                {\hbox{e.g.,~}}

\def \etal              {\hbox{et~al.~}}

\def \pc                {{\rm\ pc}}

\def \H0                {{\rm\ H_{0}}}
\def \kmsmpc            {{\rm\ km\ s^{-1}\ Mpc^{-1}}}

\def \deg               {{$^{\rm o}$}}
\def \mum               {{\rm{\mu}m}}
\def \kel               {{\rm\ K}}

\def \clumpy            {{\textsc{Clumpy}~}}
\def \clumpybb          {{\textsc{Clumpy}+Blackbody~}}
\begin{document}
  \slugcomment{Accepted in ApJ, to appear in March 2011 issue.}

  \shortauthors{Deo, Richards, Nikutta, \etal}

  \shorttitle{Dusty Tori of Luminous Type 1 Quasars}

  \title{Dusty Tori of Luminous Type 1 Quasars at $z\sim2$}

  \author{Rajesh. P. Deo\altaffilmark{1}, Gordon. T. Richards\altaffilmark{1},
    Robert Nikutta\altaffilmark{2}, Moshe Elitzur\altaffilmark{2}, Sarah C.
    Gallagher\altaffilmark{3}, {\v Z}eljko Ivezi{\'c}\altaffilmark{4},
    Dean Hines\altaffilmark{5}}

  \altaffiltext{1}{Department of Physics, Drexel University, 3141, Chestnut
    St., Philadelphia, PA 19104-2816, USA; rpd@physics.drexel.edu and
    gtr@physics.drexel.edu}

  \altaffiltext{2}{Department of Physics and Astronomy, University of
    Kentucky, Lexington, KY 40506-0055, USA}

  \altaffiltext{3}{Department of Physics and Astronomy, University of
    Western Ontario, 1151 Richmond St, PAB 213D London, ON N6A 3K7, Canada}

  \altaffiltext{4}{Department of Astronomy, University of Washington, Box
    351580, Seattle, WA 98195, USA}

  \altaffiltext{5}{Space Science Institute, Boulder, CO.}

  \begin{abstract}
    We present \textit{Spitzer} infrared spectra and ultra-violet to
    mid-infrared spectral energy distributions (SEDs) of 25 luminous type 1
    quasars at z $\sim$ 2. In general, the spectra show a bump peaking around
    3 $\mum$, and the 10 $\mum$ silicate emission feature. The 3 $\mum$
    emission is identified with hot dust emission at its sublimation
    temperature. We explore two approaches to modeling the SED: (i) using the
    \clumpy model SED from Nenkova \etal (2008a), and (ii) the \clumpy model
    SED, and an additional blackbody component to represent the 3 $\mum$
    emission. In the first case, a parameter search of $\sim 1.25$ million
    \clumpy models shows:  (i) if we ignore the UV-to-near-IR SED, models fit
    the 2--8 $\mum$ region well, but not the 10 $\mum$ feature; (ii) if we
    include the UV-to-near-IR SED in the fit, models do not fit the 2--8
    $\mum$ region. The observed 10 $\mum$ features are broader and shallower
    than those in the best-fit models in the first approach. In the second
    case, the shape of the 10 $\mum$ feature is better reproduced by the
    \clumpy models. The additional blackbody contribution in the 2--8 $\mum$
    range allows \clumpy models dominated by cooler temperatures ($T < 800
    {\rm K}$) to better fit the 8--12$\mum$ SED. A centrally concentrated
    distribution of a small number of torus clouds is required in the first
    case, while in the second case the clouds are more spread out radially.
    The temperature of the blackbody component is $\sim 1200 \kel$ as 
    expected for graphite grains.
  \end{abstract}

  \section{Introduction}

  In the unified model of active galactic nuclei (AGN)
  \citep{1993ARA&A..31..473A,1995PASP..107..803U}, the dust torus is a region
  immediately outside the accretion disk where dusty clouds are no longer
  sublimated by the radiation from the central engine. The dust torus
  reprocesses the incident ultra-violet/optical radiation from the accretion
  disk and this energy emerges in the near- and mid-infrared bands.
  \citet{2006ApJS..166..470R} presented panchromatic spectral energy
  distributions (SEDs) for 259 type 1 quasars selected from the Sloan Digital
  Sky Survey\footnote{http://www.sdss.org/}
  \citep[SDSS,][]{2000AJ....120.1579Y}. These quasar SEDs constructed from
  broad-band photometry are remarkably similar over a large range in both
  luminosity and redshift. However, \citet{2006ApJS..166..470R} noted small
  differences in the 1.3--8 $\mum$ range between optically luminous and
  optically dim quasars. \citet{2007ApJ...661...30G} investigated this
  further, and found that the 1--8 $\mum$ spectral index ($\alpha_{\nu}$) is
  strongly anti-correlated with infrared luminosity in type 1 quasars. More
  luminous quasars have bluer 1--8 $\mum$ slopes. Further, they noted a tight
  linear correlation between the ultra-violet (UV) continuum luminosity and
  the infrared luminosity for these quasars. This suggested that the observed
  near-IR emission at $3\ \mum$ in the SED of many type 1 objects is driven by
  the dust reprocessing of the intrinsic optical/ultra-violet continuum from
  the accretion disk, as had been noted previously
  \citep{1969Natur.223..788R,1979ApJ...230...79N,1986ApJ...308...59E,1987ApJ...320..537B,1989ApJ...347...29S}.
  As a recent example, the near-IR emission is clearly visible in the spectrum
  of Mrk~1239 \citep{2006MNRAS.367L..57R}.
  
  Theoretical work on the response of accretion disks to radiation and
  hydromagnetic pressure suggests that outflow of matter is associated with
  all accretion disks in the form of a wind coming off the surface of the disk
  \citep{1994ApJ...434..446K,1995ApJ...454L.105M,2000ApJ...543..686P}. The
  dusty torus itself may be the outermost part of this accretion disk wind
  close to the equator of the system
  \citep{1994ApJ...434..446K,2006ApJ...648L.101E}. Disk-winds have a natural
  dependence on luminosity through radiation pressure, and this begs the
  question: ``is the structure of the dusty torus related to the physics of
  the accretion disk?''. The need for proper radiation transfer treatment of
  clumps in dusty tori was recognized in the pioneering early studies
  \citep{1988ApJ...329..702K,1992ApJ...401...99P,1995MNRAS.272..737R}, and was
  fully developed by \citet{2002ApJ...570L...9N}. More recently,
  \citet{2008ApJ...685..147N} presented their model in detail (denoted by
  \textsc{Clumpy} hereafter).
  
  Significant effort has been invested in understanding the torus dust
  distributions with various groups favoring both clumpy and smooth dust
  density distributions
  \citep{2002ApJ...570L...9N,2005A&A...436...47D,2005A&A...437..861S,2006MNRAS.366..767F,2006A&A...452..459H,2008A&A...482...67S,2008ApJ...685..147N}.
  The primary difference between clumpy and smooth models is that of the dust
  temperature distributions \citep[see Fig.~3 of][]{2008A&A...482...67S}.
  While in smooth density models, the temperature steadily declines with
  radius from the inner wall, clumpy models can show a range of temperatures
  at different distances from the central source. This effect occurs primarily
  due to the shadowing effect from the finite size of clouds. The inner faces
  of clouds are directly exposed to radiation from the central source, and are
  hence hotter, while their outer faces are much cooler. And because of
  clumpiness, clouds farther out in radius can still have their inner faces
  exposed directly to radiation from the central source.
  
  The effective optical depth in a clumpy torus is a function of the number
  density of clouds in the central regions of the torus. This important model
  construction has resulted in better fits to both low-resolution
  \textit{Spitzer} spectra \citep{2009ApJ...705..298M,2009ApJ...707.1550N},
  and high-resolution interferometric observations of dust tori in NGC 1068
  \citep{2004Natur.429...47J} and Circinus \citep{2007A&A...474..837T}.
  
  \clumpy models appear to be the most promising set of models with a wide
  range of applications to both active galactic nuclei (AGN)
  \citep[e.g.,][]{2006ApJ...640..612M} and merger-driven ultra-luminous
  infrared galaxies (ULIRG) \citep{2007ApJ...654L..45L}. Other notable models
  that employ clumps arranged in a disk-like geometry include
  \citet{2008A&A...482...67S} and \citet{2006A&A...452..459H}. For example,
  \citet{2008ApJ...675..960P} employed clumpy torus models from
  \citet{2006A&A...452..459H} to fit their optically obscured but
  infrared-bright sources at high-redshift.
  
  \clumpy models show changes in their near-IR continua based on the average
  number of clouds ($N_{0}$) encountered along a radial equatorial ray
  \citep[see Fig.~6 in][]{2008ApJ...685..160N}. Using \textit{Spitzer} mid-IR
  spectroscopy of high redshift quasars it is then possible to constrain the
  parameters of their dusty tori. While \textit{Spitzer} archives are rich in
  observations of low-redshift Seyfert galaxies, they are deficient in
  high-redshift observations of radio-quiet quasars at the peak of the quasar
  activity in the Universe. In this paper, we present such observations as
  obtained with the Infrared Spectrograph (IRS) on board \textit{Spitzer}. Our
  goals include: (1) presenting high-quality mid-IR quasar spectra covering
  rest-frame 2--12$\mum$ for comparison to the low-redshift templates already
  available
  \citep[e.g.,][]{2005ApJ...625L..75H,2005ApJ...633..706W,2006AJ....132..401B,2006ApJ...640..579G,2006ApJ...653..127S,2006ApJ...649...79S};
  (2) testing the validity of \clumpy torus models by fitting the observed
  spectra with model SEDs. Using good-quality IRS spectra we hope to model the
  10 $\mum$ region properly and constrain \clumpy torus parameters for these
  luminous quasars.
  
  The properties of the sample and reduction process of the IRS spectra are
  presented in Section~\ref{sec:data}. The IRS spectra and SEDs of the sample
  are discussed in Section~\ref{sec:the-spectra}. \clumpy torus models are
  summarized in Section~\ref{sec:clumpy-torus-models}, and
  Section~\ref{sec:model-fits} presents results of model fits to ultra-violet
  to mid-IR SEDs. Results are summarized in Section~\ref{sec:summary}. In all
  calculations, we assume a standard cosmology with $H_{0} = 71 \kmsmpc$,
  $\Omega_{\rm{M}} = 0.27$, and $\Omega_{\rm{Vac}} = 0.730$.
  
  \section{Data}
  \label{sec:data}
  
  \subsection{The Sample}
  \label{sec:the-sample}
  
  Our primary sample includes those quasars from \citet{2006ApJS..166..470R}
  that are 1) in the 1.6--2.2 redshift range, 2) not BAL quasars, and 3)
  require IRS exposure times less than 2 hours to achieve S/N $\sim 15$ in
  each of the four IRS low resolution bandpasses. There are 25 such objects in
  the \citet{2006ApJS..166..470R} sample. Four of these have already been
  targeted by IRS (Program 3046, PI: I. Perez-Fournon). Most objects from this
  sample also have \textit{Spitzer} InfraRed Array Camera
  \citep[IRAC,][]{2004ApJS..154...10F} observations from the \textit{Spitzer}
  Wide-area InfraRed Extragalactic (SWIRE) survey \citep{2003PASP..115..897L}.
  
  The redshift range 1.6--2.2 was chosen to provide rich diagnostics in both
  the optical and ultra-violet via SDSS spectroscopy and photometry, and in
  the mid-IR range via \textit{Spitzer} observations. At these redshifts, the
  SDSS spectroscopy samples the crucial 1000--3500\AA~range giving a direct
  measurement of the strength and shape of the ultra-violet (UV) continuum.
  The 1.6--2.2 redshift range allows the rest-frame 2--14 $\mum$ range to be
  redshifted into the IRS low-resolution bandpass of 5.2 to 38 $\mum$. The
  IRAC bandpasses (at 3.6, 4.5, 5.8, and 8.0$\mum$) provide coverage of the
  rest-frame 1--3 $\mum$, thus sampling rest frame 1--14$\mum$. The model
  torus SED changes significantly in this region depending on the average
  number of clouds along the line of sight, their average temperatures and
  radial distributions \citep[see Fig.~6 in][]{2008ApJ...685..160N}.
  
  The objects chosen are listed in Table~\ref{tab:sample}, along with a
  summary of the low-resolution spectroscopic observations. The
  \textit{Spitzer} IRS low-resolution data comes mainly from programs 50087
  (PI: G.T. Richards, 16 objects), 50328 (PI: S.C. Gallagher, 5 objects) and 4
  archival datasets from program 3046 (PI: I. Perez-Fournon) as mention above.
  Out of the 16 objects for which observations were requested in program
  50087, we were able to obtain observations of 15 objects and 1 observation
  (SDSSJ163021) failed to a peak-up lock on a nearby bright star-forming
  galaxy instead of the quasar. Only this source does not have an IRS
  spectrum, but we use its SED for analysis. Model fits for this source are
  unreliable due to lack of IRS spectrum. All 5 objects from program 50328
  were observed. Table~\ref{tab:continua} provides the photometric
  measurements as obtained from the SDSS DR7 catalogue along with absolute
  $i$-band magnitude and $\Delta(g - i)$ values \citep[see figure~5
  of][]{2003AJ....126.1131R}. The redshifts in Table~\ref{tab:continua} are
  taken from updated SDSS redshift catalog provided by
  \citet{2010MNRAS.405.2302H}. Table~\ref{tab:continua2} provides the 2MASS,
  IRAC and MIPS measurements from the 2MASS and SWIRE databases.
  Table~\ref{tab:continua3} provides continuum measurements from the reduced
  IRS spectra at 3, 5, 8 and 10 $\mum$ in the rest-frame.

  \subsection{Data Reduction}

  We obtained the basic calibrated data (BCD) products processed with the
  standard \textit{Spitzer} IRS pipeline (version S18.7.0) from the
  \textit{Spitzer} Science Center (SSC) archive. We cleaned the BCD images
  using the IRSCLEAN software package to fix rogue pixels using SSC supplied
  masks, and a weak thresholding of the pixel histogram. We co-added the
  multiple data collection event (DCE) image files into one image for each
  module, spectral order and the ``nod'' position (\eg SL, 1st order, 1st
  nod-position) using the fair-coadd option in SMART. We differenced the
  co-added images from the opposite ``nod'' positions to remove the sky
  background. The spectra were extracted using the optimal extraction option
  within the SMART package. All the image combining and spectrum extraction
  operations were carried out using SMART \citep{2004PASP..116..975H}. We also
  checked our extractions using the SPICE program. We obtained average S/N of
  $\sim$6--10 for 4 archival spectra from program 3046, $\sim$ 10--13 for
  spectra from program 50087, and $\sim$ 25--35 for the spectra from proposal
  50328. These S/N estimates were commensurate with the pre-determined
  configuration of each observation. Figure~\ref{fig:seds} displays the
  observed spectra plotted along with SEDs.

  \section{Spectra and SEDs}
  \label{sec:the-spectra}
  
\begin{figure*}[ht]
    \begin{center}
      \includegraphics[width=\textwidth]{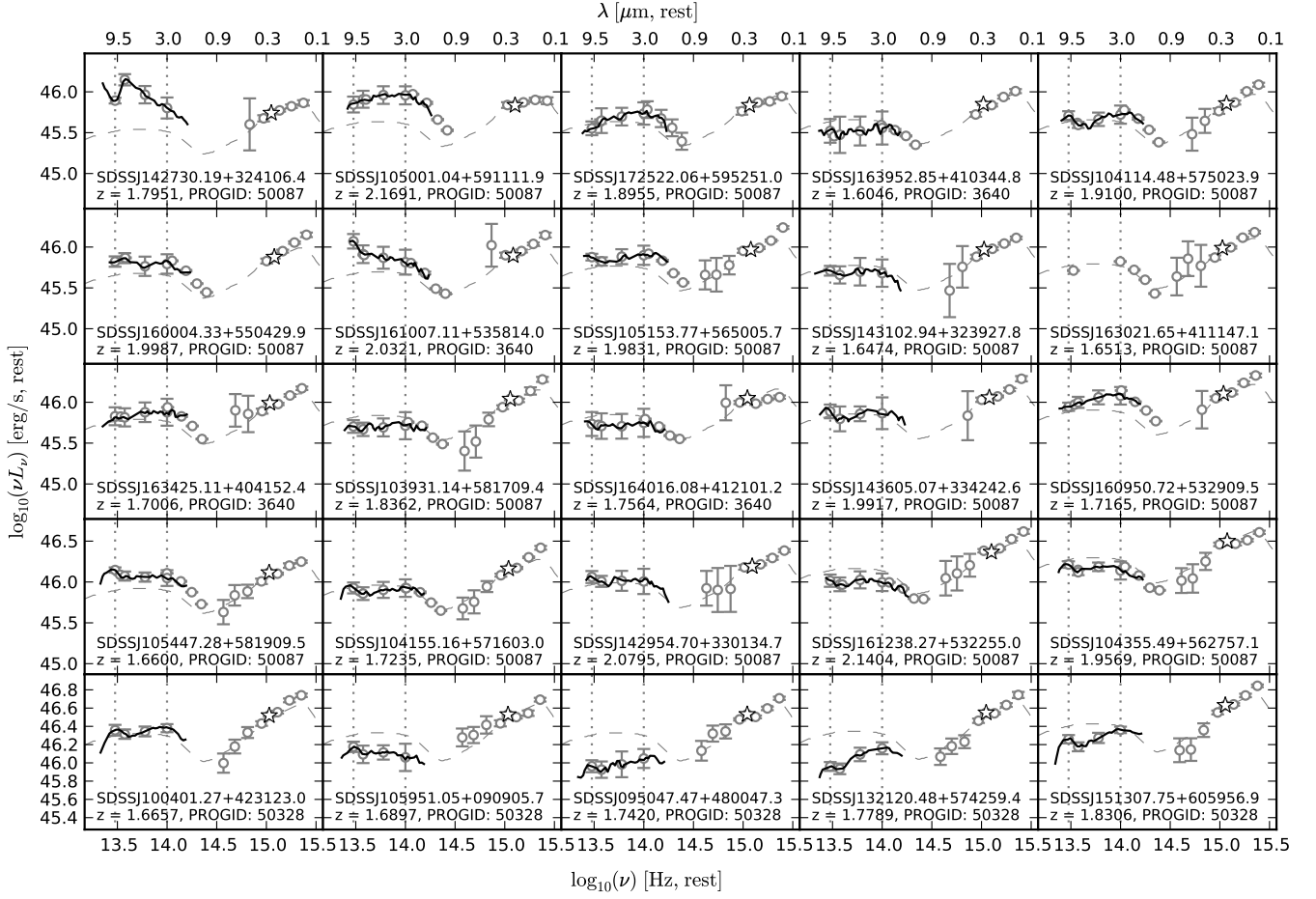}
    \end{center}
    \caption{Spectral energy distributions from ultra-violet to mid-infrared
      for the $z\sim2$ quasars in our sample. The continuum in the optical/UV
      region is sampled by the SDSS photometry; near-IR/optical is sampled by
      the 2MASS photometry; \textit{Spitzer}/IRAC photometry samples the
      downturn of the near-IR bump around $\sim 1.6\mum$; Photometric
      measurements on \textit{Spitzer}/IRS spectra provides continuum points
      at 3, 5, 8 and 10 $\mum$. \textit{Spitzer}/IRS mid-infrared spectra
      smoothed by a 35-point polynomial are shown by the thick dark line. The
      S/N ratio of the IRS spectrum varies from 10--13 for objects from
      program 50087, and it is 25--35 for objects from program 50328. The S/N
      for archival objects from program 3640 is between 6--10. The 3 and 10
      $\mum$ regions where the graphite dust blackbody, and the silicate
      features peak, respectively, are indicated with vertical dotted lines.
      For comparison, we have over-plotted the mean quasar SED from
      \citet{2006ApJS..166..470R} as a gray dashed line. The mean SED is
      normalized to the SDSS $i$ band luminosity density. The $i$-band is
      indicated by an open star symbol. Objects are sorted by their $i$-band
      luminosities, beginning with least-luminous in top-left corner to most
      luminous in bottom-right corner.}
    \label{fig:seds}
  \end{figure*}

  In general, the spectra show two features peaking at $\sim3$ and $\sim10\
  \mum$ (see Figure~\ref{fig:seds}, features are marked by vertical dashed
  lines) in $\nu L_{\nu}$ units. The infrared spectral index ($\alpha_{\nu}$)
  from 3--8 $\mum$ ranges from -0.49 to -1.82, with a median of -0.86. The 10
  $\mum$ emission feature is the well-known 10 $\mum$ silicate emission
  feature due to the Si-O stretching mode of the silicate molecule. This
  emission feature was well-known in stellar spectra for a long time
  \citep[\eg][]{1988ApJ...333..305L}, but has only recently been detected in
  quasar spectra
  \citep{2005AN....326R.556S,2005ApJ...625L..75H,2005ApJ...629L..21S} due to
  the sensitive spectroscopy and broad wavelength coverage possible with
  \textit{Spitzer}.
  
  The weakness of the 10 $\mum$ emission feature in IR spectra of local type 1
  AGN had motivated suggestions of presence of different chemical compositions
  and/or size distributions of dust grains
  \citep{1993ApJ...402..441L,2001A&A...365...37M}. Instead \clumpy models of
  \citep{2002ApJ...570L...9N} make use of the clumpy nature of the dusty
  medium to improve model fits to the 10 $\mum$ region. However, as we will
  see ahead, different sublimation temperatures and radii for graphite and
  silicate grains remain an important issue to be resolved in torus models.
  
  The emission peaking between 2 and 4 $\mum$ can be attributed to the
  blackbody emission from dust close to its sublimation temperature
  \citep{1969Natur.223..788R, 1979RvMP...51..715D, 1987ApJ...320..537B}, which
  is typically expected to be $T\gtrsim1500\kel$ for graphite dust. This hot
  dust emission has long been expected based on broad-band IR data
  \citep{1989ApJ...347...29S}. Measurement of the strength of this feature
  relative to longer wavelength mid-IR emission is important because it can
  give constraints on the inclination of the torus assuming a disk-like
  configuration \citep{1993ApJ...418..673P,2000ApJ...528..179M}. Recent
  advances in near-IR ground-based spectroscopy has lead to observations of
  the near-IR bump in Mrk 1239 \citep{2006MNRAS.367L..57R} and NGC 4151
  \citep{2009ApJ...698.1767R}.
  
  
  Figure~\ref{fig:seds} shows the SEDs constructed using the photometric data
  points from Tables~\ref{tab:continua}, \ref{tab:continua2}, and
  \ref{tab:continua3}. Also over-plotted for each object is the IRS spectrum
  along with the mean quasar SED template from \citet{2006ApJS..166..470R}
  scaled to the SDSS $i$-band luminosity for each object. While the mean SED
  captures the overall trend quite well, individual spectra reveal significant
  differences from the mean SED. Objects with similar UV luminosities can have
  different relative IR power ($\sim$ 0.3 dex). Obscured sources (for \eg
  SDSSJ142730, program 50087) are significantly more IR luminous than sources
  with similar observed UV luminosities (for \eg SDSSJ172522, program 50087)
  that are probably not as strongly obscured based on their optical SDSS
  spectra. This trend is reflected in the mean SEDs constructed by
  \citet{2006ApJS..166..470R}.
  
  \section{CLUMPY Torus Models}
  \label{sec:clumpy-torus-models}
  
  We use the \clumpy torus models from \citet{2008ApJ...685..147N} to fit the
  complete SEDs. The models are constructed by assuming an intrinsic AGN SED
  that heats the dust clouds \citep[see Figure~4 of][]{2008ApJ...685..147N}.
  We do not considered the effects of a different intrinsic AGN SED. This
  effect was partially studied by \citet{2008ApJ...685..147N} (see their
  Fig.~12), and it is expected that the SED longward of 1 micron should not
  change significantly. \clumpy models contain a standard Galactic mix of
  silicates (53\%) and graphite (47\%) dust grains. We have not explored
  changes in composition, and size distribution of dust grains \citep[see
  \eg][]{1993ApJ...402..441L}, and contributions from species other than
  silicates \citep[\eg][]{2007ApJ...668L.107M}. These areas should be
  addressed by future work on torus models.
  
  The \clumpy torus model is realized as a collection of individual molecular
  clumps/clouds arranged in a toroidal structure around the central accretion
  disk. In reality, this region is likely to be a continuous extension of the
  outer accretion disk \citep{2006ApJ...648L.101E}. The primary parameters of
  the \clumpy torus model are described below.
  \begin{enumerate}
  \item \textbf{$N_{0}$}: It is the \textit{average} number of clouds along a
    radial equatorial ray in a given model. It represents the normalization of
    a Gaussian distribution of clouds around the equatorial plane. The total
    number of clouds intersecting a given equatorial ray is different for
    different lines of sight. The intrinsic AGN continuum can escape along
    many different lines of sight, and the observed mid-infrared 10 $\mum$
    silicate features can be seen in emission even for lines of sight close to
    the equator. The total \textit{effective} optical depth to the continuum
    source is thus a function of the number of clouds along the line of sight
    and optical depth of each cloud.
  \item \textbf{$\tau_{V}$}: Each of the clouds/clumps are assumed to have the
    same optical depth $\tau_{V}$ in the V-band. Assuming standard Milky Way
    dust extinction with $R_{V} = 3.1$, $A_{V}/\tau_{9.7\mum} = 18.0 \pm 1.0 $
    \citep{1984MNRAS.208..481R,2003dge..conf.....W}, and $A_{V} =
    1.086\,\tau_{V}$; only when $\tau_{9.7\mum} \ge 1 $ or $\tau_{V} \gtrsim
    16.5$, we will notice the effects of self-absorption on the 10 $\mum$
    feature.
  \item \textbf{$Y$}: The radial extent of the torus $Y$, which is the ratio
    of the outer ($R_{o}$) to the inner radius ($R_{d}$) of the torus. The
    inner radius depends on the onset of dust sublimation due to the incident
    UV radiation from the accretion disk \citep{1987ApJ...320..537B}. See also
    Eq.~1 in \citet{2008ApJ...685..160N}. The radial extent $Y$ of the torus
    decides the infrared turn-over at long wavelengths ($\lambda \gtrsim 30\
    \mum$).
  \item \textbf{$q$}: The clouds are distributed along the radius with a
    power-law distribution ($r^{-q}$) parametrized with the exponent ``$q$''.
    For $q > 1$, the clumps are concentrated closer to $R_{d}$. When the
    clumps are packed closer to $R_{d}$, the resultant infrared SED is
    dominated by the emission from dust close to its sublimation temperature,
    and there is little long-wavelength mid-IR emission. The corresponding
    width of the SED \citep{1993ApJ...418..673P} in this case is also small.
  \item \textbf{$\sigma$}: The torus angular width $\sigma$, is the width
    of the Gaussian distribution of clumps around the equatorial plane. Thick
    tori (large $\sigma$) generate redder 3--8 $\mum$ continua (in
    $\lambda F_{\lambda}$ units).
  \item \textbf{$i$}: The models produce the infrared SED longward of $\sim
    1~\mum$ for each inclination $i$ from $0^\circ$ (face-on) to $90^\circ$
    (edge-on) in steps of $10^\circ$.
  \end{enumerate}
  The torus models are constructed using the radiative transfer code, \clumpy
  \citep{2002ApJ...570L...9N}. The tabulated SEDs for different parameters are
  accessible from the \textsc{Clumpy} project
  website\footnote{\url{http://www.pa.uky.edu/clumpy}}.
  
  \clumpy dust density distributions differ from smooth distributions in one
  important aspect: in smooth dust distributions the temperature is uniquely
  determined by the distance from the source of radiation. While this is also
  roughly true for clumpy distributions, the presence of lines of sight with
  different dust columns allows both hot and cold temperature regions to
  co-exist at similar radial distances. This leads to a greater dependence of
  the output SED on $N_{0}$, $\tau_{V}$ and $q$. The primary motivating factor
  for considering clumpy models for the torus is interferometric observations
  of local AGN \citep{2004Natur.429...47J} which constrain the tori to be
  physically small ($R_{o} \lesssim \rm{a\ few}\pc$).
  
  \section{Model Fits}
  \label{sec:model-fits}
  
  To fit our data with the \clumpy torus models, we adopt the procedure
  developed by \citet{2009ApJ...707.1550N}. We analyze the distributions of
  best fitting \clumpy torus parameters for each quasar in our sample. We
  consider the following grid of parameters,
  \begin{itemize}
  \item $q$ = 0.0--3.0, in steps of 0.5
  \item $N_{0}$ = 1--15, in steps of 1
  \item $\tau_{V}$ = 5, 10, 20, 30, 40, 60, 80, 100, 150
  \item $Y$ = 5, 10, 20, 30, 40, 50, 60, 70, 80, 90, 100
  \item $\sigma$ = 15--70, in steps of 5
  \item $i$ = 0--90, in steps of 10
  \end{itemize}
  in all, $\sim 1.25$ million possible combinations of model parameters. Very
  large values of $Y$ and $\tau_{V}$, present in the original model grid in
  \citet{2009ApJ...707.1550N}, are excluded here as the objects under study
  are type 1 quasars; with, in most cases silicate 10$\mum$ feature in
  emission.
 
  Each model is scaled and fitted such that the overall fitting error $E$ is
  minimized. We adopt Eq.~1 from \citet{2009ApJ...707.1550N} shown below.
  \begin{equation}
    E = \frac{1}{N}\sqrt{\sum^{N}_{i=1}\left(\frac{ F_{AGN} \cdot f_{i}^{m} - f_{i}^{\rm{obs}}}{\sigma_{i}} \right)}
  \end{equation}
  
  Here, $f_{i}^{\rm{obs}} = \lambda_{i} F_{i}$, are the observed SED data
  points that are interpolated at model grid points denoted by $\lambda_{i}$
  (see ahead for why we take this approach); $f_{i}^{m}$ are the corresponding
  model SED points; $\sigma_{i}$ are the 1-$\sigma$ errors on $\lambda_{i}
  F_{i}$. The scaling of the model, $F_{AGN}$, provides a measure of the
  infrared luminosity of the \clumpy torus, which can be converted into an
  estimate of the bolometric luminosity of the system.
  
  
  For each parameter, we construct a discrete distribution of values by
  selecting a sample of well-fit models. For each model, the fitting error $E$
  is computed from Eq~1. The model with minimum value of fitting error,
  $E_{\rm Min}$, is considered to be the best-fit model. Further, a relative
  error, $E_{r} = 100 \times \left|E - E_{\rm Min}\right|/E_{\rm Min}$, for
  each model is constructed. Models that differ by 10\% from the minimum value
  $E_{\rm Min}$ are considered to represent the distribution of parameter
  values that best represents the data for a given quasar. For each parameter,
  we consider the mode of the distribution of parameter values as the most
  probable value of the parameter for a given quasar. Note that the
  best-fitting value may not be the most probable one. The 90\% confidence
  intervals for a parameter are also computed.
  
  The model SEDs are scaled, and fitted to the data SEDs constructed from the
  photometric data, and the IRS spectrum. We attempt the fitting procedure for
  all 1.25 million model SEDs, and record their respective relative error
  $E_{r}$. Parameter distributions are then constructed where the acceptance
  criteria to form the samples are $E_{r} \sim$ 10\%, 20\% and 30\%. We find
  that the distributions gradually become flatter or uniform as the relative
  error criterion is relaxed. Thus, a narrower distribution suggests a better
  constrained parameter value.
  
  To measure how well a parameter is constrained we use the discrete
  Kullback-Leibler divergence. The KL divergence measures the similarity of
  two histograms (or discrete distributions) of identical sampling $k$. The KL
  divergence is written as,
  \begin{equation}
    D_{KL} = \sum_{k} Po_{k} * \log_{2}(Po_{k}/Pr_{k}) / \log_{2}(N) \nonumber
  \end{equation}
  with $N$ the number of sampled bins, $Pr$ the prior distribution of
  parameter values (uniform in this analysis), and $Po$ the posterior
  distribution of parameter values (histogram of ``accepted'' parameter
  values). The normalization $\log_{2}(N)$ ensures that $D_{KL} = 1.0$ when
  all accepted models happen to have a parameter value within a single bin. A
  $D_{KL}$ value close to 1 indicates a better constrained parameter.
  
\begin{figure*}[!ht]
    \begin{center}
      \includegraphics[height=\textwidth]{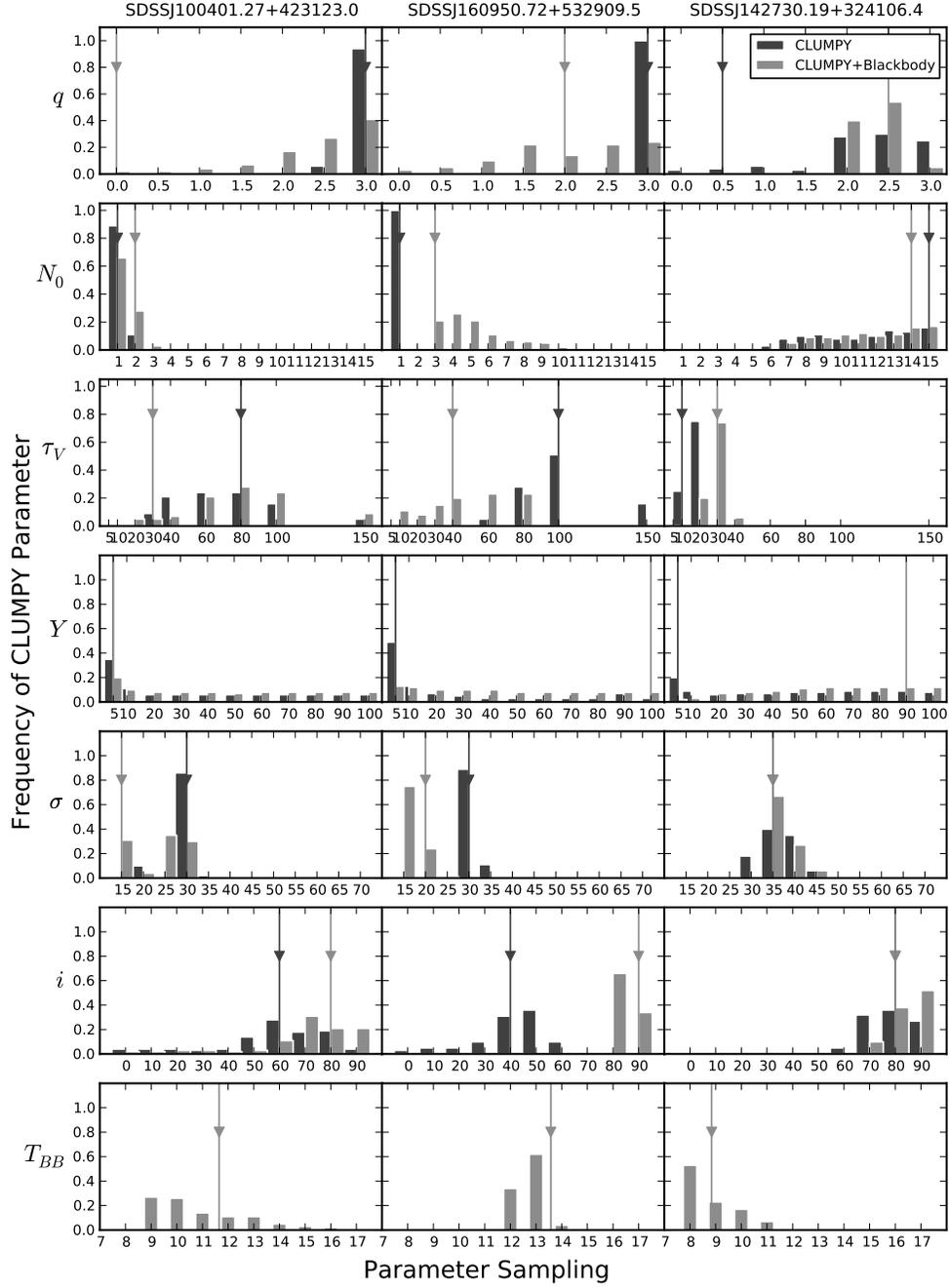}
    \end{center}
    \caption{Distributions of model parameters for three representative
      sources, SDSSJ100401.27+423123.0, SDSSJ160950.72+532909.5, and
      SDSSJ142730.19+324106.4, are shown. The distributions are formed from
      models whose relative error is within 10\% of the minimum fitting error
      of the best fit model. The vertical lines with arrow-heads show the
      parameter value for the best-fitting model, which is often close to the
      mode of the parameter distribution for well constrained parameters. Note
      that radial extent of the torus $Y$, is unconstrained (almost flat
      distributions), while most other parameters are well constrained. It is
      interesting that the IR SED of these type 1 quasars (by sample choice)
      requires high inclination tori, which we think is an artifact of the
      two-component fitting. The temperature sampling as shown ($T_{BB}$) is
      in units of 100 K.}
    \label{fig:pardist}
  \end{figure*}

  Figure~\ref{fig:pardist} shows the parameter distributions corresponding to
  three sources from our sample for brevity. For each parameter there are
  three figures from left to right corresponding to the three sources:
  SDSSJ100401.27+423123.0, SDSSJ160950.72+532909.5, and
  SDSSJ142730.19+324106.4. SDSS100401 shows strong near-IR emission, and has
  high UV luminosity (Figure~\ref{fig:seds}). SDSSJ160950 is weak in the UV
  and also has weak 3 and 10 $\mum$ features; SDSS142730 has a deep 10 $\mum$
  absorption feature. This source shows a power-law optical/UV continuum in
  its SDSS spectrum, but its emission lines are absorbed (and may be a
  BALQSO), which is consistent with its mid-IR nature. For each source, we
  show the distribution of \clumpy parameter values that forms by accepting
  models that fall within 10\% of the best-fit model.
  
  It should be noted that we are fitting the entire SED from UV to MIR, and
  the IRS range is more densely sampled than the photometry. To avoid problems
  due to uneven spectral sampling, the data SED was resampled to the
  wavelength grid of the models. This is not a significant issue since we are
  interested in fits to the broad-band features of the SED such as the optical
  AGN power-law, the near-IR bump and the 10 $\mum$ feature. The \clumpy model
  SED is better sampled near 9.7 $\mum$ than elsewhere, thus, this improves
  the fit to the 10 $\mum$ region without biasing the fit to be weighted more
  by the 2--8 $\mum$ continuum. Another important point to be noted is that
  the selection of the model (AGN+TORUS) SED is also constrained by the
  optical/UV portion of the data SED. While we do not investigate changes in
  the intrinsic AGN SED, by including fits to the optical SDSS photometry, we
  are preferentially selecting model SEDs that satisfy consistent flux density
  scaling in both UV and mid-IR regime at the same time.
  
  \subsection{Model fits using the \clumpy SED}
  
  Initially, we used only the \clumpy model SED to fit the data SED. The
  best-fit values of the parameters for each model are given in
  Table~\ref{tab:torus_best_fit_without_bb}. Example model fits are shown in
  the left-hand panels of Figure~\ref{fig:clumpy_fits}. The best-fitting
  models of the entire sample have $N_{0} \sim 1$, $\tau_{V} \sim $ 20--100,
  $q \gtrsim 2$, $\sigma \sim 15$, and $i \sim $ 60--80 (see
  Table~\ref{tab:torus_best_fit_without_bb} and dark bars in
  Figure~\ref{fig:pardist}). The radial extent $Y$ of the torus is
  unconstrained with parameter distributions nearly flat over the sampling
  grid.
  
  We find that models with $N_{0} \sim 1$, $q \sim 3$, $\tau_{V} \lesssim 10$,
  $Y \sim 5$, and $\sigma \sim 15$ show peaked 10 $\mum$ silicate emission
  features for all values of the inclination of the line of sight. For
  $\tau_{V} \lesssim 15$, all wavelengths longer than $\sim 1.5\,\mum$ have
  $\tau < 1$, and the dust emission is optically thin
  \citep{2008ApJ...685..160N}. In this case, the SED simply follows the shape
  of the dust absorption co-efficients, which decreases rapidly at longer
  wavelengths in the mid-IR. The observed spectra should then have blue 3--8
  $\mum$ continua, which is indeed the case for luminous objects like
  SDSS100401 and SDSSJ151307 (both from program 50328), as can be seen in
  Figure~\ref{fig:seds}, bottom row of panels.
  
  Further, $q$ is well-constrained in the case of single-component models to a
  high value of 2--3 in the case of most objects. This suggests a steep radial
  distribution of clumps, with most clumps concentrated close to $R_{d}$.
  \citet{2008ApJ...685..147N} show that whenever $q \gtrsim 2$, $Y$ is
  fundamentally unconstrained. As most clumps are closer to $R_{d}$ in this
  case, the absolute size of the torus does not matter; the output SEDs from
  tori of all sizes look the same. On the other hand, for sources with $q \sim
  0$, the clump distribution is flatter/spatially extended, and $Y$ can be
  constrained much better for such sources as cooler temperatures contribute
  at longer wavelengths.
  
  Increasing $N_{0}$, $\tau_{V}$, and/or $Y$ causes the SED to become redder
  in the 2--8 $\mum$ wavelength range, and the overall flux density peak
  shifts to longer mid-infrared wavelengths (due to the Wien displacement
  law). The increasing $N_{0}$ and $\tau_{V}$ essentially increases the
  obscuration due to the torus, and leads to increased contribution from the
  cooler parts of the clouds. This effect can be seen by comparing best-fit
  values of $N_{0}$ for SDSS142730 (Table~\ref{tab:torus_best_fit_without_bb})
  with the rest of the sample. Larger $N_{0}$ at smaller $\tau_{V}$ and small
  $Y$ apparently still produce deep absorption features. Larger $Y$ has
  similar effect if $q \lesssim 1$, as clouds are more spread out radially,
  and hence cooler. Thus, detecting a blue SED in the 2--8 $\mum$ range
  suggests small $N_{0}$, $\tau_{V}$, and $Y$, along with a radially steep
  distribution ($q \gtrsim 1$) of clouds. This conclusion however comes with a
  caveat: while it is clear that the near-IR emission is generated by the dust
  close to its sublimation point, the strong silicate emission features
  predicted by the \clumpy models with these parameter configurations are not
  observed.
  
  The near-IR emission is fitted well by \clumpy models with $N_{0} \sim 1$,
  $\tau_{V} \sim 5$ and $q > 1$, the 10 $\mum$ feature profiles are not
  well-fit by the same models. The model 10 $\mum$ profiles are more peaked
  than observed profiles, which are broad and shallow. We note that this
  uncertainty about the origin of the near-IR emission in torus models was
  also encountered previously in the study by \citet{1993ApJ...418..673P},
  where they also had to employ an additive blackbody component to represent
  the near-IR contribution separate from their mid-IR torus component. Even in
  smooth density models, where dust temperatures are functions of radial
  distance from the source, use of a common sublimation temperature for
  graphite and silicate dust leads to this effect. Using different sublimation
  radii for different grain populations is computationally expensive, which
  could explain some of these discrepancies.
  
  Fitting UV/optical continuum and mid-IR together highlights the need for an
  additional blackbody component (see left panels of
  Figure~\ref{fig:clumpy_fits}). \citet[][see their
  Figure~1]{2008ApJ...675..960P} also came to similar conclusions in their
  effort to fit high-z extremely obscured sources with clumpy torus models
  from \citet{2006A&A...452..459H}. This appears to be a common problem to all
  clumpy models constructed so far.
  
  \subsection{Additive Near-IR Blackbody Emission}
  
  To improve fits to the 10 $\mum$ features, we considered a linear
  combination of a blackbody and a \clumpy SED (hereafter \clumpybb model) as
  explored also by \citet{2009ApJ...705..298M} for PG quasars. The best-fit
  values of the parameters for this model are given in
  Table~\ref{tab:torus_best_fit_with_bb}. The model fits are shown in the
  middle panels of Figure~\ref{fig:clumpy_fits}.
  
  The additive blackbody component represents emission from the very hot dust
  at the inner edge of the torus. \clumpy models use standard Galactic dust
  composition consisting of both silicates (53\%) and graphite (47\%). The
  blackbody emission around $3\ \mum$ is expected to be a result of emission
  from graphite grains. As we saw in the last section, this emission can be
  matched using \clumpy models with $N_{0} = 1$, $q > 1$, $\tau_{V} \sim $
  5--10. The problem is not matching the near-IR blackbody emission, but
  matching the 10 micron emission using the same model parameters. The
  silicate emission in these models is stronger than observed in the spectra.
  This is likely to be an artifact of constructing a single dust grain type
  that is a linear combination of individual grain emission efficiencies. This
  approach is taken in \textsc{DUSTY} \citep{1999astro.ph.10475I}, the
  underlying radiation transfer code for \clumpy. Requiring a fit to only the
  10 micron region selects models with weak emission at 3 microns. Additional
  blackbody contribution above that obtained from the Clumpy models possibly
  indicates the presence of an extended graphite zone, where silicates are
  depleted, something that is not accounted for by \clumpy models assuming a
  single composite grain type at all radii. This extended graphite zone may
  have a smooth density profile.
  
  The \clumpybb models provide better fits to the 10 $\mum$ feature (see
  panels on the right in Figure~\ref{fig:clumpy_fits}). A much larger range of
  model parameters becomes accessible (see Table~\ref{tab:torus_par_stats})
  due to the addition of the hot blackbody component. However, this process
  also weakens any constraints that could be placed on $N_{0}$, $\tau_{V}$ and
  $q$ as a larger number of models are now accepted by the relative error
  criterion. Thus, an additive blackbody is but a temporary stop-gap, until
  the models are expanded. Since the additive blackbody is ad-hoc, the
  resulting total model SED is phenomenological in nature.
  
  Overall, the \clumpybb models prefer more extended tori ($q \sim 1$, $Y \sim
  $ 50--100) with a somewhat larger number of clouds ($N_{0} \sim $ 5--15) of
  large optical depths ($\tau_{V} \sim $ 40--150) and \textit{high
    inclinations} $i \sim 80$\deg. Our sources are selected to be type 1
  objects, and we expect the inclination of our line of sight to be smaller
  than $\sim 60^\circ$. In the case of \clumpybb model, $q$ appears to be
  constrained only for source SDSSJ142730, which has a deep 10 $\mum$
  absorption feature. The torus angular width $\sigma$ is relatively better
  constrained in the \clumpybb models than in the single-component model (See
  Table~\ref{tab:torus_par_stats}).
  
  
  The median ratio of integrated flux ($\lambda F_{\lambda}$ longward of 1
  micron) between the blackbody and the \clumpy model is $0.19\pm0.11$ for our
  objects. In most luminous objects, this ratio is about $0.15$, which
  suggests that the very hot dust emits a small portion of the $L_{\rm IR}$
  \citep[see also][]{1993ApJ...418..673P}, and the bulk of the emission occurs
  in the ``warm'' 8--25 $\mum$ part of the torus, and this part also likely
  contains most of the dust mass because the dust emissivity decreases at
  longer wavelengths. 
  
  It is interesting to note that in the case of the \clumpybb model the
  $D_{KL}$ values are not close to 1 for all parameters, which suggests that
  multi-component fits weaken the constraints the near-IR data put on torus
  model parameters. Adding a blackbody component makes constraining \clumpy
  torus parameters difficult without additional far-IR data. Observations
  using the \textit{Herschel} space observatory will likely provide a measure
  of the contribution of the torus against that contributed by circum-nuclear
  star formation \citep[][see their Figure~6]{2007ApJ...666..806N}, and allow
  better constraints to be put on the torus models in the long-wavelength
  regime.
  
  \section{Observed Silicate Features}
  \label{sec:silicate-features}
  
  The 10 $\mum$ silicate emission feature gets broader and weaker with
  increasing $\tau_{V}$, $N_{0}$, $\sigma$, and $Y$. None of our objects show
  peaked 10 $\mum$ silicate emission profiles in the spectra, indicating that
  hot dust generating the near-IR emission is depleted in silicate dust, and
  that the 10$\mum$ region receives contribution from multiple ``colder than
  sublimation temperature'' sources which likely make the feature broader and
  weaker. Right-hand panels in Figure~\ref{fig:clumpy_fits} show the fits of
  silicate features in the presence of an extra blackbody component.

  In most objects, the feature either peaks around 9.7 $\mum$ (SDSS100401) or
  is mostly flat (SDSS160950). In some cases, there is a well-defined plateau
  from 9.7 to 11.4 $\mum$ (SDSS151307, last right-most panel in
  Figure~\ref{fig:clumpy_fits}). We find that with the \clumpybb fits, the
  models mostly reproduce the observed shapes within the errors of the
  observations, with the exception of emission around 11.3 $\mum$. This
  suggests presence of dust species other than silicates in these quasar
  spectra \citep[see
  also][]{2005ApJ...625L..75H,2005ApJ...629L..21S,2007ApJ...668L.107M}. We
  find that with the exception of excess flux around 11 $\mum$, the silicate
  features in 14 out of 25 sources are fitted well.
  
  Another issue in fits to the 10 $\mum$ features is the observed shift of the
  feature peak in quasar spectra \citep[see \eg Fig~3
  of][]{2005ApJ...625L..75H}. If this shift is a real effect is still
  uncertain, however we note that radiation transfer in clumpy media as
  demonstrated by the fits in this paper may explain the varied shapes and
  apparent shift of the feature peak.

  \section{Summary}
  \label{sec:summary}

  We present \textit{Spitzer}/Infrared Spectrograph (IRS) observations of a
  sample of optically luminous type 1 quasars at z$\sim$2. Their rest-frame
  2--12 $\mum$ infrared spectra show two prominent features peaking at $\sim$
  3 and 10 $\mum$. The 10 $\mum$ feature is the 10 $\mum$ silicate emission
  feature, commonly observed in \textit{Spitzer} observations of other type 1
  AGN \citep{2005ApJ...625L..75H,2005AN....326R.556S,2005ApJ...629L..21S}. The
  3 $\mum$ bump is the expected signature of the hottest thermal dust emission
  from the inner region of the dust torus. There is a strong correlation
  between the optical/UV and infrared luminosities
  \citep{2007ApJ...661...30G}, and the detection of this near-IR bump in a
  sample of optically luminous high redshift quasars, shows that the
  optical/UV continuum heats the dust in the inner torus, which then radiates
  in the thermal near- to mid-infrared.

\begin{figure*}[!ht]
    \begin{center}
      \includegraphics[width=0.32\textwidth]{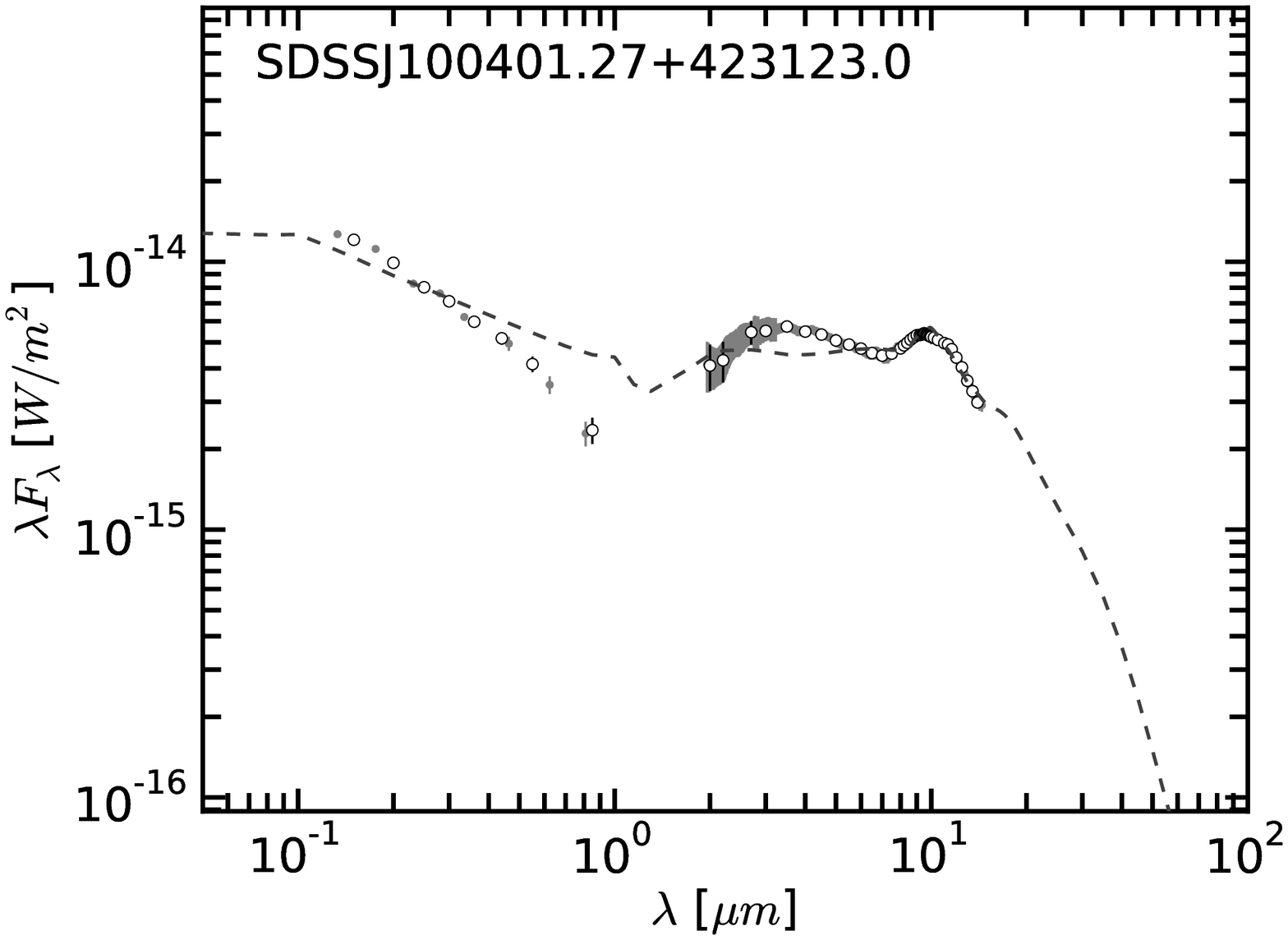}%
      \includegraphics[width=0.32\textwidth]{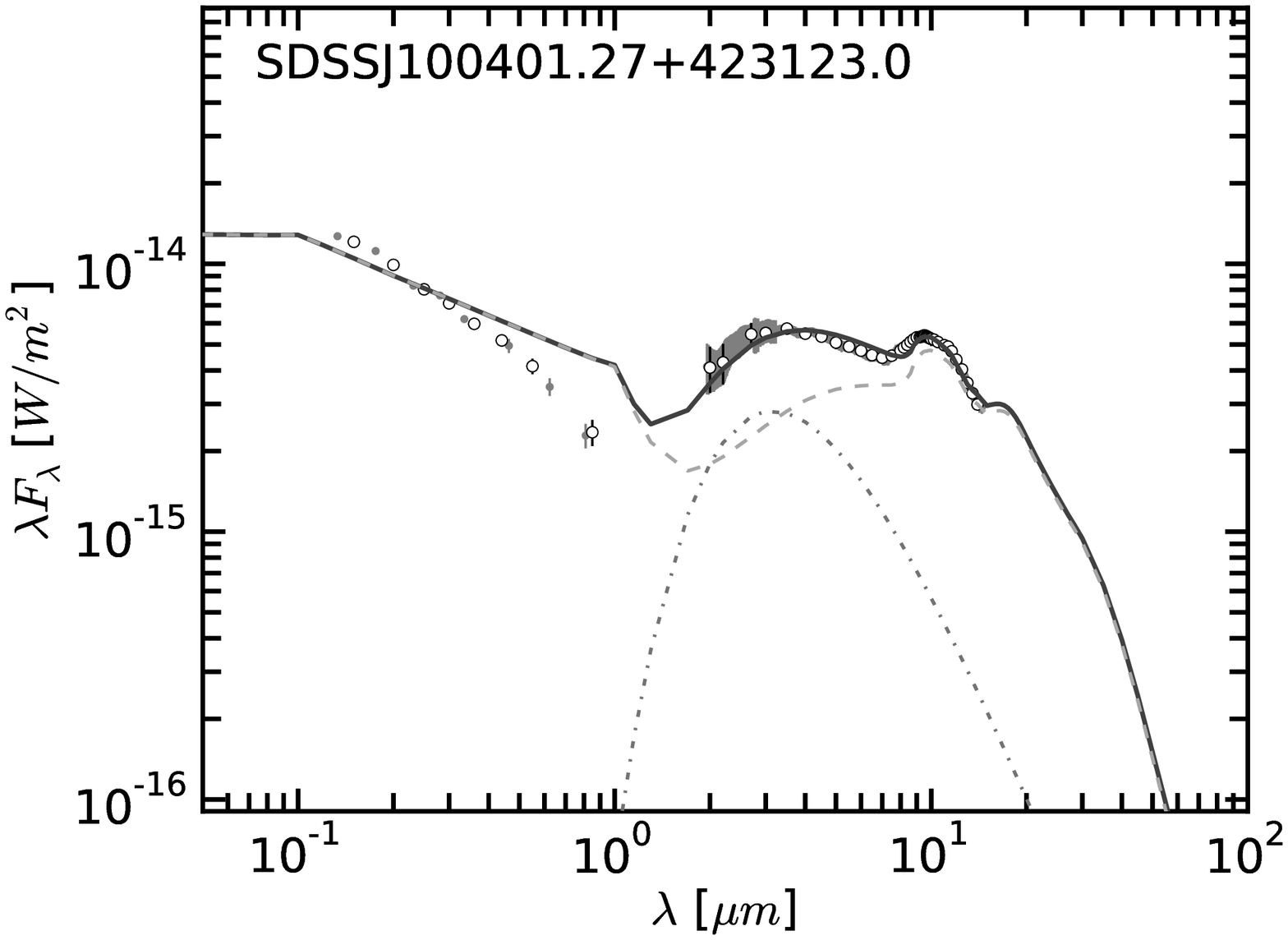}%
      \includegraphics[width=0.32\textwidth]{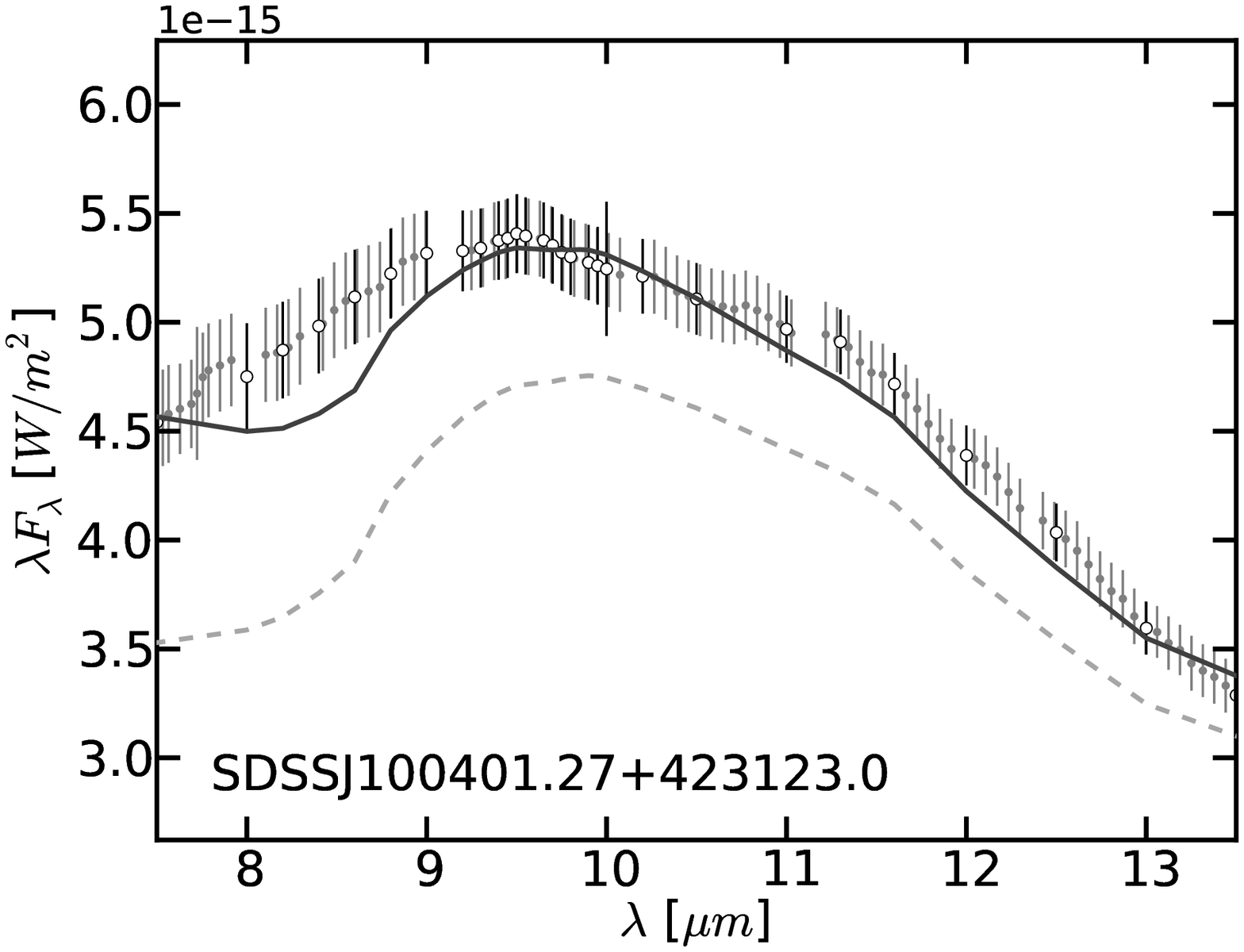}
      \includegraphics[width=0.32\textwidth]{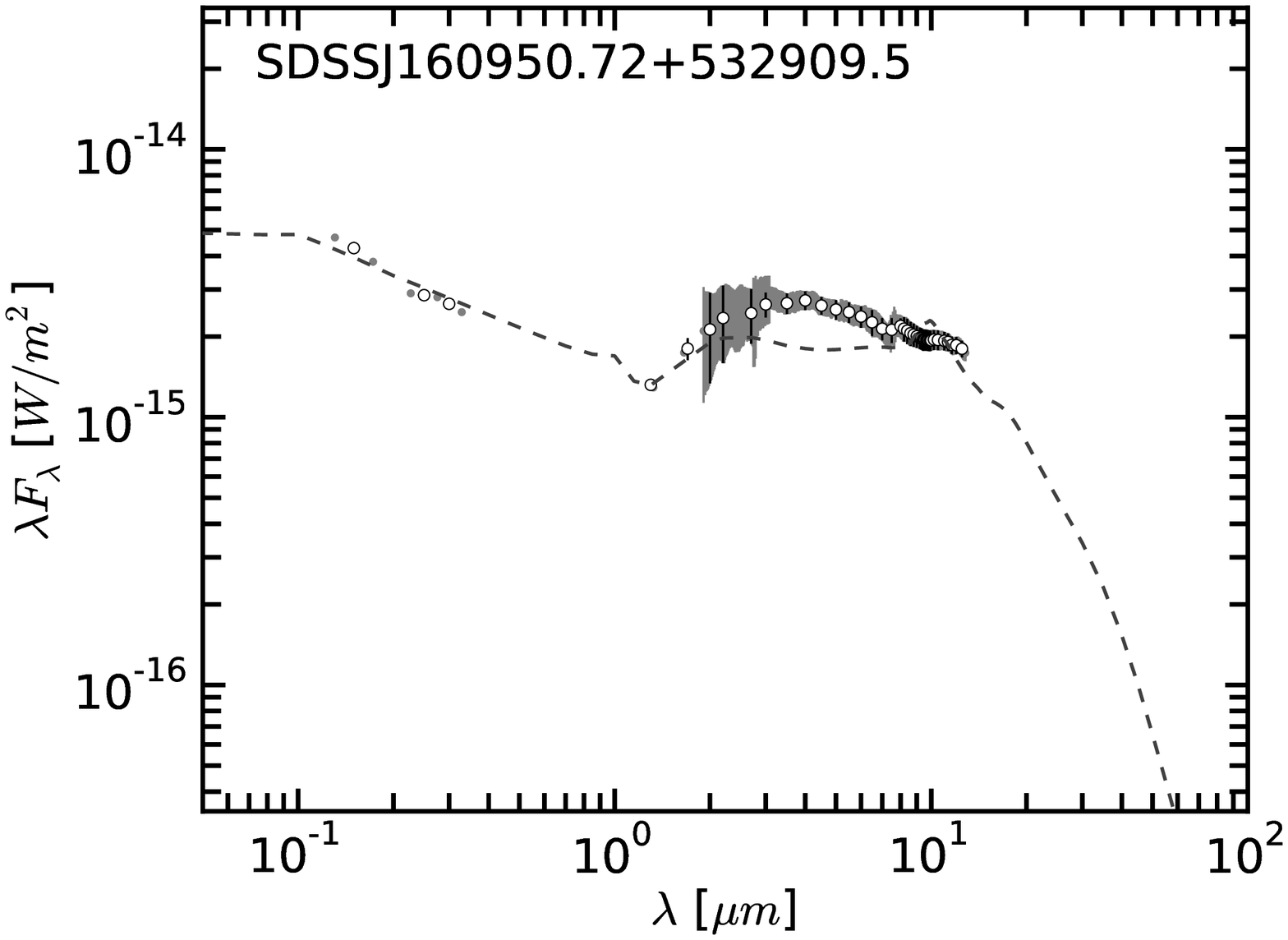}%
      \includegraphics[width=0.32\textwidth]{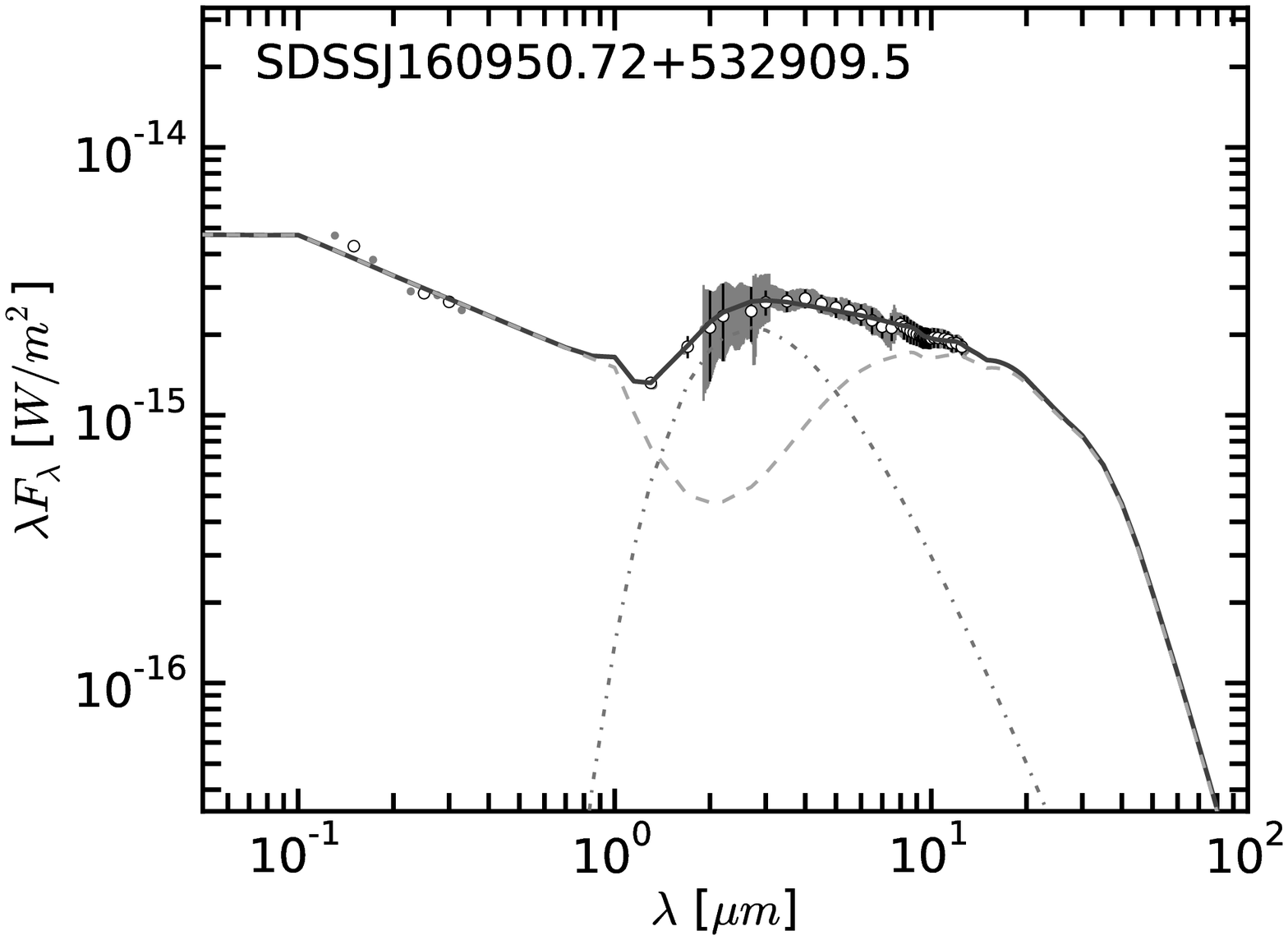}%
      \includegraphics[width=0.32\textwidth]{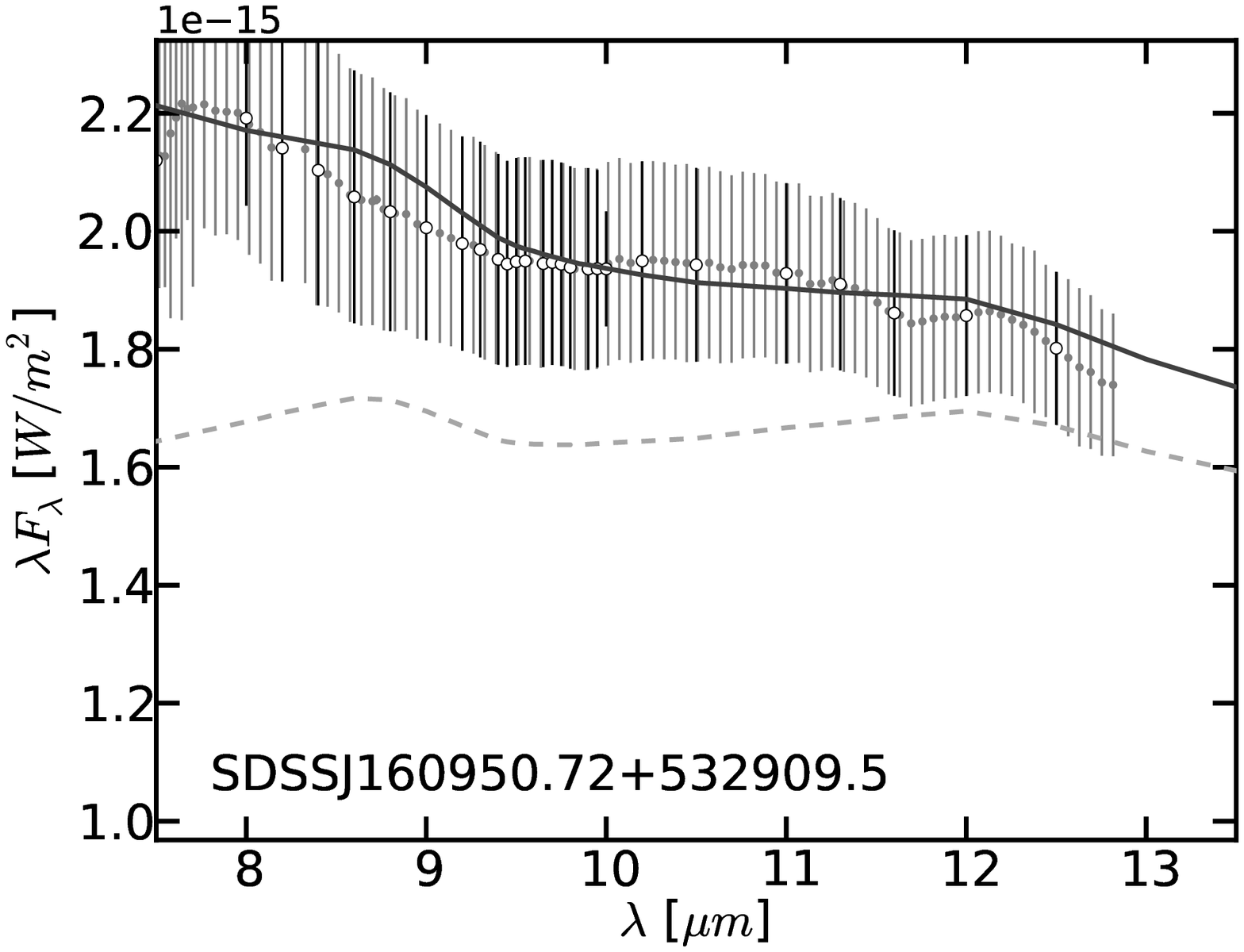}
      \includegraphics[width=0.32\textwidth]{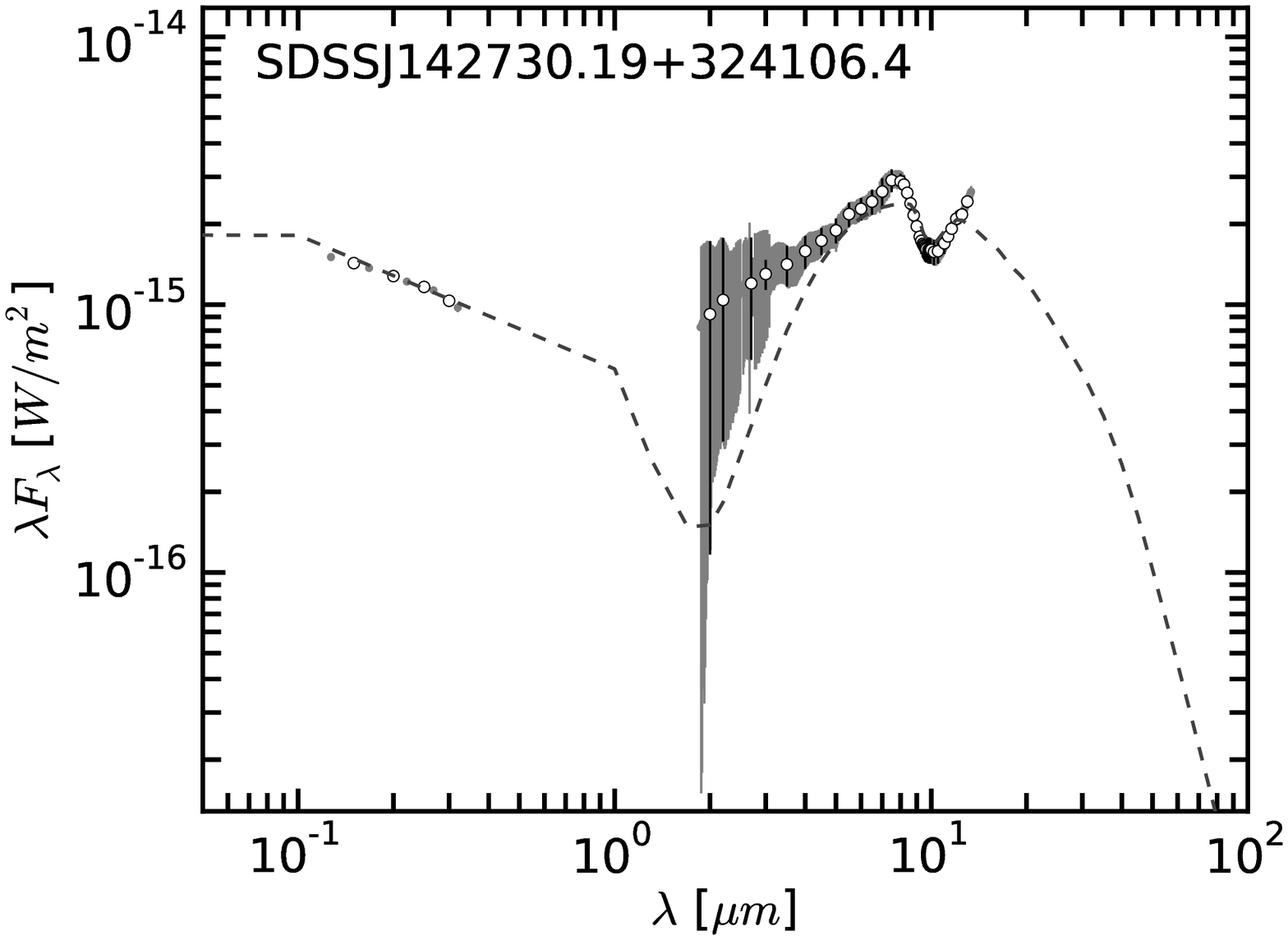}%
      \includegraphics[width=0.32\textwidth]{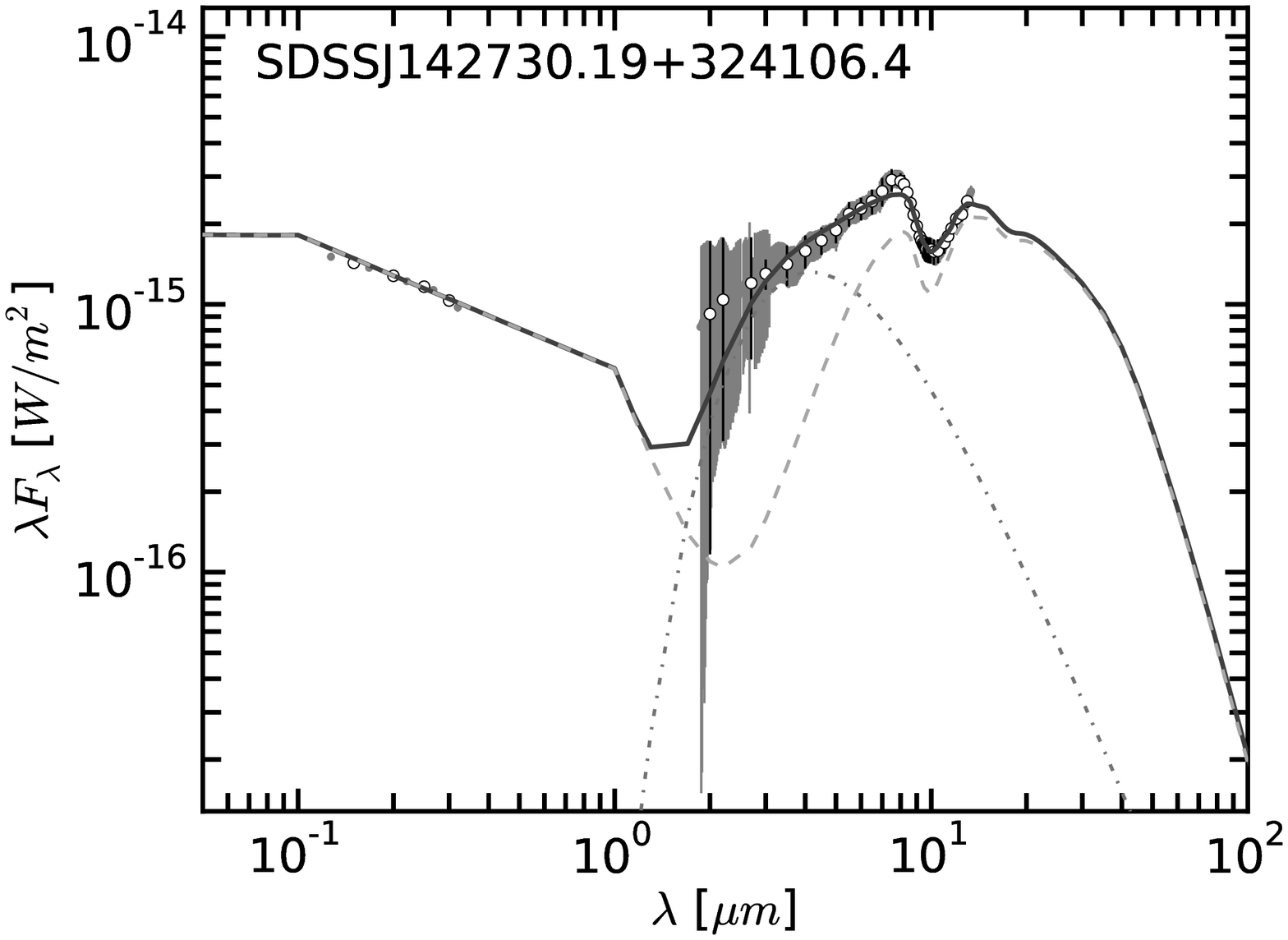}%
      \includegraphics[width=0.32\textwidth]{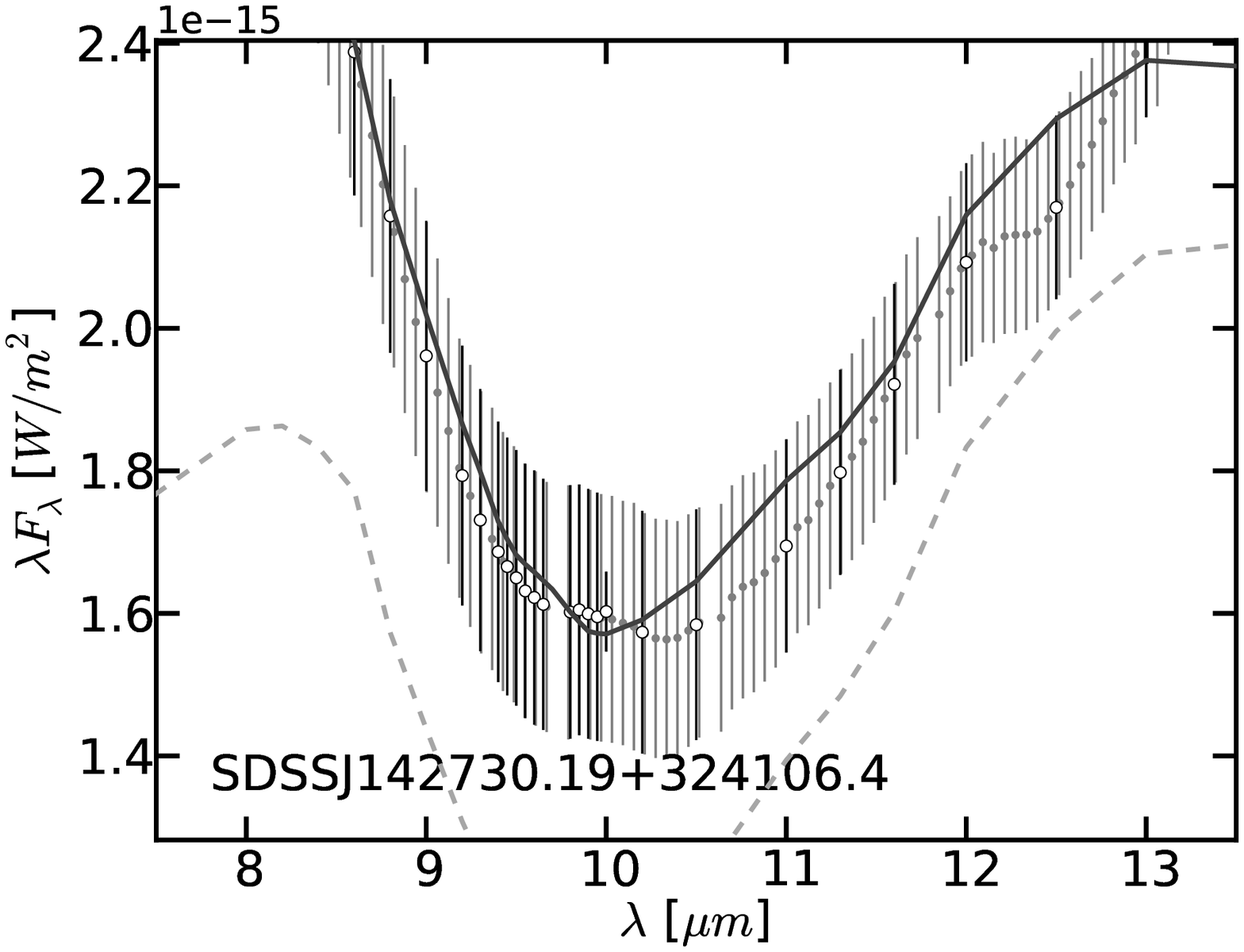}
      \includegraphics[width=0.32\textwidth]{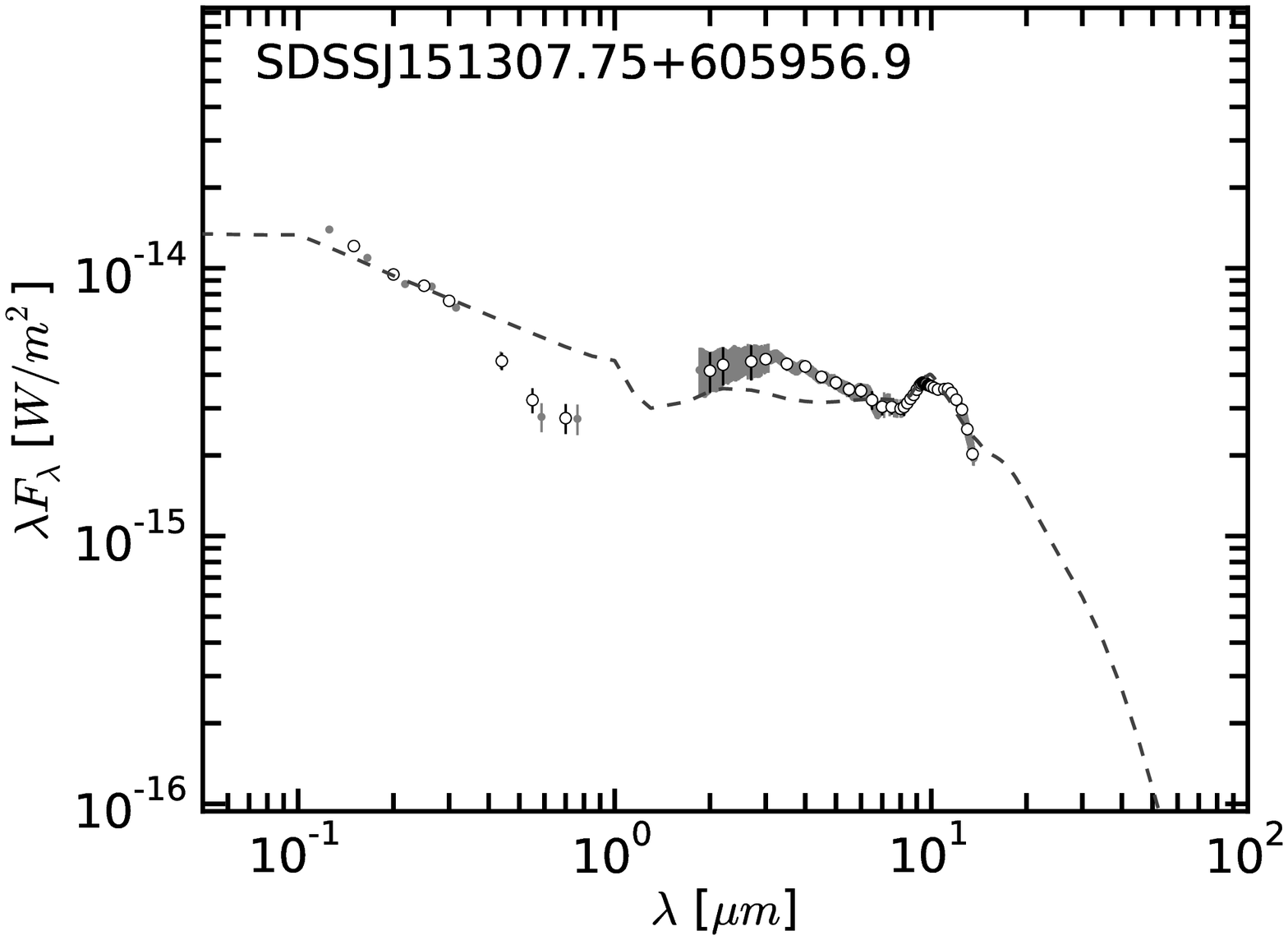}%
      \includegraphics[width=0.32\textwidth]{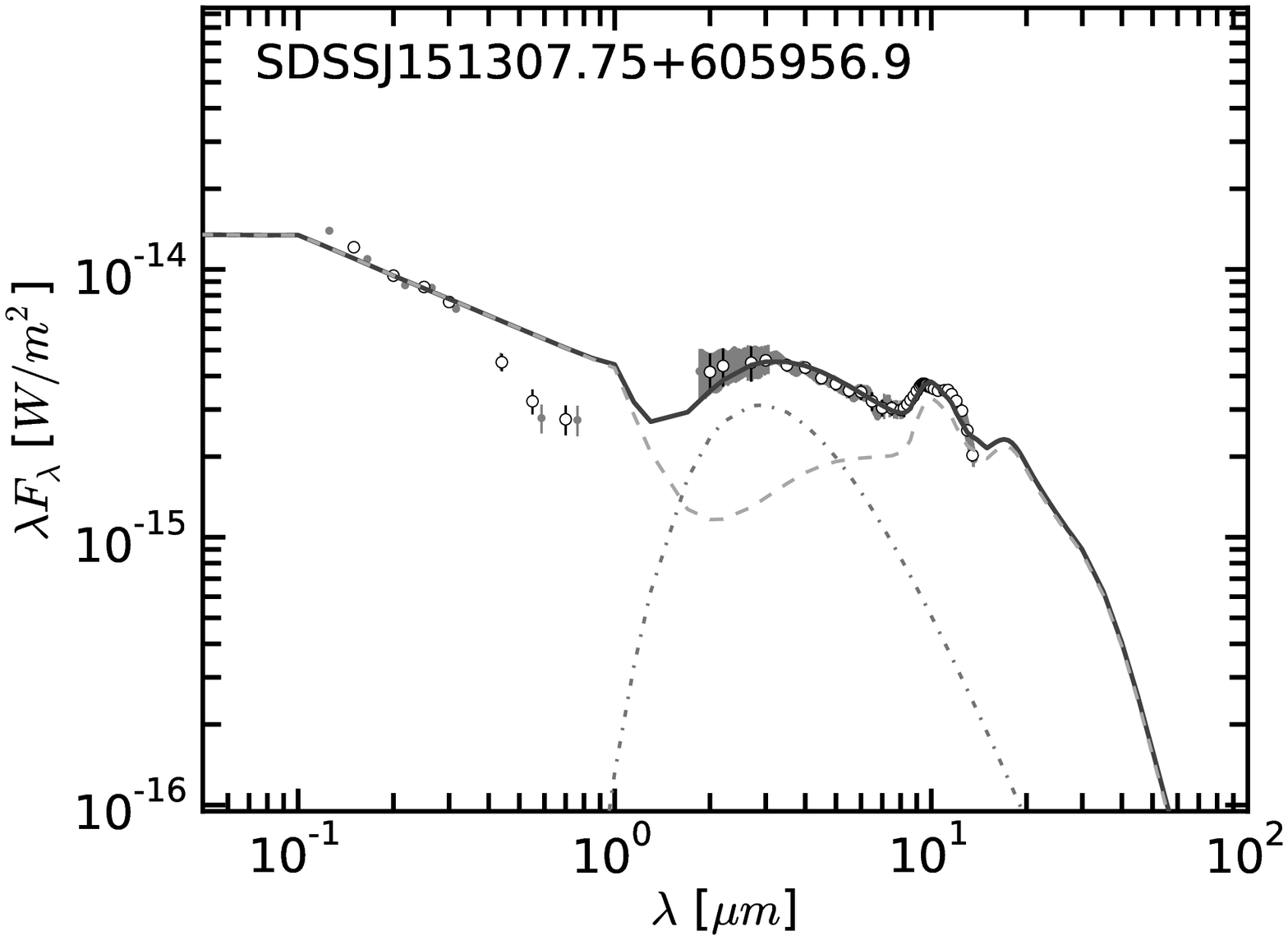}%
      \includegraphics[width=0.32\textwidth]{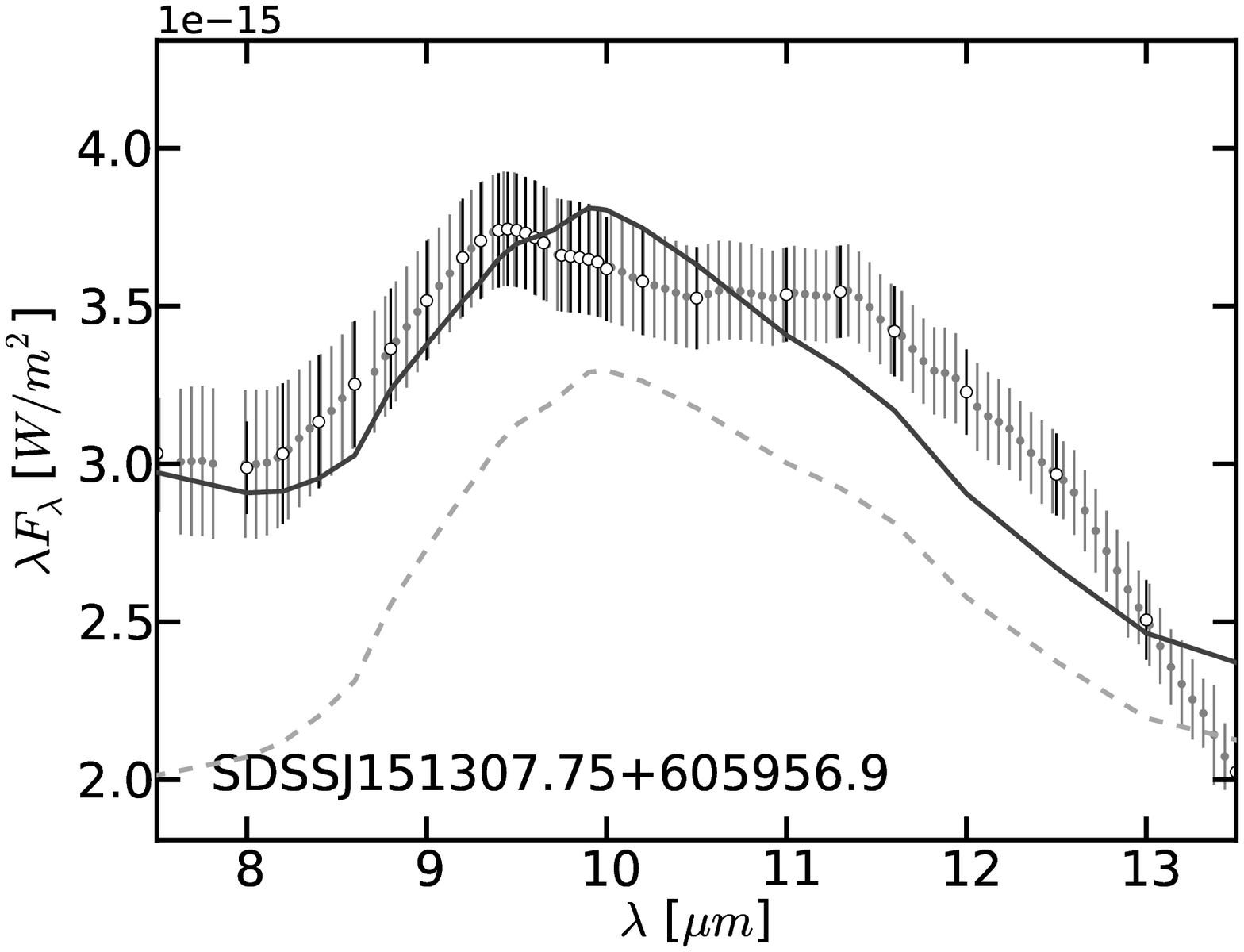}
    \end{center}
    \caption{Best-fitting \clumpy torus models (dashed lines) overlaid on the
      resampled (open circles with error bars), and original data SEDs (filled
      gray circles with error bars) and the IRS spectra (clustered gray dots
      with error bars). Model fits for four sources are shown,
      SDSSJ100401.27+423123.0, SDSSJ160950.72+532909.5, and
      SDSSJ142730.19+324106.4, SDSSJ151307.75+605956.9. Panels on the left
      show fits using only the \clumpy model (dashed line). Note the excess 3
      $\mum$ emission. Middle panels show fits using a \clumpy model (light
      gray dashed line), and a hot blackbody component (dash-dotted line) to
      represent the excess 3 $\mum$ emission. The overall fit (dark solid
      line) improves by incorporating the blackbody component. The panels on
      the right show a blow up of the 10 $\mum$ region from the middle panel.
      The 10 $\mum$ region of SDSSJ151307.75+605956.9 (bottom row, right
      panel) shows two peaks, one around 9.7 and one around 11.3 $\mum$
      possibly indicative of crystalline dust species. See
      Table~\ref{tab:torus_best_fit_with_bb} for values of best fitting
      parameters.}
    \label{fig:clumpy_fits}
  \end{figure*}

  We fit the spectra and the UV-to-MIR SED with \clumpy torus models
  \citep{2008ApJ...685..147N}. This is the first time such fits have been
  attempted to spectroscopically confirmed high-z quasars with near-IR data.
  We considered two different approaches. In the first case, we use the
  \clumpy model SED. These \clumpy torus models provide good fits to the 2--8
  $\mum$ part of the spectrum, if we only fit data longward of 1 $\mum$.
  Models with average number clouds along a radial equatorial ray ($N_{0}$)
  $\sim 1$, optical depth through each cloud ($\tau_{V}$) $\lesssim 10$, and a
  radial distribution of clouds ($r^{-q}$) described by a power-law exponent
  ($q$) $\sim 3$ fit IRS spectra (not complete SEDs) with a strong hot-dust
  bump very well. The $q \sim 3$ values suggests that the hot dust component
  is more centrally concentrated as expected. However, the 10 $\mum$ silicate
  emission features of these models show strongly peaked profiles, and the 10
  $\mum$ feature in the observed spectra are more broad and flat. This problem
  can be partially removed by fitting the entire SED from UV-to-MIR; using
  this long lever-arm, the \clumpy model SED is consistently weaker than the
  observed SED in the 1--7 $\mum$ range (see left panels of
  Figure~\ref{fig:clumpy_fits}), highlighting the lack of additional near-IR
  contribution in the models, if both UV and IR data is fitted together.
  
  To accurately model the 10 $\mum$ silicate emission features, and remove the
  above inconsistency, we considered the \clumpybb model where we fit the
  spectra and the SED with a linear combination of a hot dust blackbody and a
  \clumpy model. In these fits, the clumpy models provide good fits to the 10
  $\mum$ region, while the blackbody contributes more strongly to the region
  between 2--8 $\mum$. Use of the additional blackbody leads to a stronger
  contribution of the \clumpy model to the far-IR emission. Whether this is a
  real effect may be tested using far-IR facilities like \textit{Herschel}.
  
  We compared the infrared properties of this sample to the low-redshift PG
  quasar sample ($z\sim0.1$) from the \textit{Spitzer} archives, and find that
  the primary difference in the 2--8 $\mum$ range between low- and high
  redshift samples is the absolute luminosity. There are however significant
  object-to-object differences in the 10 $\mum$ silicate emission features,
  which point to real differences in the dust structure of their tori. In few
  cases, such as SDSSJ142945, the 9.7$\mum$ peak of the silicate feature
  appears shifted to longer wavelengths. Just as other observations have noted
  the presence of different dust species
  \citep{2005ApJ...625L..75H,2005ApJ...629L..21S,2007ApJ...668L.107M}, we note
  a feature around ~11.3 $\micron$ in some sources that may be due to
  crystalline silicates \citep{2007ApJ...668L.107M}.
  
  The 10 $\mum$ feature shapes in 14 out of 25 objects are well-reproduced by
  \clumpy models, the agreement is weak in other cases mostly due to lack of a
  clear emission feature. Presence of additional dust species also seems to
  contribute to this issue. More work is necessary to connect the near-IR
  emission with the rest of the torus structure. The lack of near-IR
  contribution in the torus models with clumpy media (in general) appears to
  be rooted in not considering the balance of amounts of silicate and graphite
  grains as a function of distance from the source.
  
  However, we find that the near-to-mid IR SED analysis is a powerful tool to
  distinguish between different distributions of $q$, $N_{0}$ and $\tau_{V}$
  in \clumpy models. Observing a blue 3--8 $\mum$ continuum indicates that the
  source is compact ($q > 1$) with $N_{0} \sim 1$. A redder continuum may
  require a more extended ($q < 1$) distribution of clumps with $N_{0} \sim
  10$ and $\tau_{V} \sim 30$. Further, improvements in fits using the complete
  UV-to-MIR SED suggests the importance of using UV/optical data if available.
  Further FIR data where the contribution from cold dust associated with star
  formation in the host galaxy of the quasar may be dominant
  \citep{2007ApJ...666..806N}, is also important. The radial extent of the
  torus ($Y$) is constrained by the location of the FIR turn-over in the
  infrared SED; however contribution from cold dust in the host galaxy is also
  dominant in the same region, disentangling these contributions will be
  interesting \citep[see for example][]{2010A&A...518L..33H}.
  
  In a \clumpy torus, the probability of viewing the AGN as a type 1 object
  depends more strongly on $N_{0}$ and $\tau_{V}$, than on the inclination to
  the line of sight $i$. Using multi-component models decreases this
  sensitivity of the model SED to parameters like $N_{0}$. This is observed in
  the number of accepted models in Table~\ref{tab:torus_par_stats}; even for
  objects with S/N $\sim 25$ (SDSS100401, SDSS151307), the number of accepted
  models is $\lesssim 1000$. The argument in favor of \clumpybb models is that
  they represent the complete data range better, and adding a blackbody
  component improves the fits to the 10 $\mum$ region (right panels in
  Figure~\ref{fig:clumpy_fits}), even in case of objects like SDSS142730 that
  should be dominated by the \clumpy model alone.
  
  Addition of the blackbody component to represent the near-IR emission does
  not by itself represent a failure of \clumpy models, but suggests that more
  detailed treatment of the origin of the near-IR emission is required. The
  composite grain approximation assumed in radiative transfer calculations
  \citep[DUSTY][]{1999astro.ph.10475I} may lead to stronger 10 $\mum$ features
  than would be generated in the actual dust sublimation transition region.
  This effect is also seen in models of \citet{2005A&A...437..861S} that use
  the standard MRN dust grain mixture, and obtain strong 10 $\mum$ emission
  features in their SEDs. As the models fits in this paper show, \clumpy
  models can reproduce the 10 $\mum$ shapes adequately. Differences in number
  density of dust grains of different sizes and compositions with distance
  from the continuum source likely contribute to the nature of near-IR
  emission. This dust sublimation region may also be spread out over an
  extended region rather than in a thin AGN-facing layer of the cloud as
  assumed in \clumpy models. Future clumpy torus models should consider both
  these effects to properly model the near- to mid-IR SEDs of active galaxies.
  
  \acknowledgments
  
  We thank the anonymous referee for their comments that significantly
  improved this paper. This work is based on observations and archival data
  from the Spitzer Space Telescope, which is operated by the Jet Propulsion
  Laboratory, California Institute of Technology under a contract with NASA.
  Support for this work was provided by NASA through awards (RSA's 1353801 and
  1365236) issued by JPL/Caltech. GTR was supported in part by an Alfred P.
  Sloan Research Fellowship. SCG thanks the National Science and Engineering
  Research Council of Canada for support. The IRS was a collaborative venture
  between Cornell University and Ball Aerospace Corporation funded by NASA
  through the Jet Propulsion Laboratory and Ames Research Center. SMART was
  developed at Cornell University and is available through the Spitzer Science
  Center at Caltech. This research has also made use of NASA's Astrophysics
  Data System Bibliographic Services.
  
  {\it Facilities:} \facility{Spitzer}
  
  \bibliography{ms}

\begin{thebibliography}{59}

\bibitem[{{Antonucci}(1993)}]{1993ARA&A..31..473A}
{Antonucci}, R. 1993, \araa, 31, 473

\bibitem[{{Barvainis}(1987)}]{1987ApJ...320..537B}
{Barvainis}, R. 1987, \apj, 320, 537

\bibitem[{{Buchanan} {et~al.}(2006){Buchanan}, {Gallimore}, {O'Dea}, {Baum},  {Axon}, {Robinson}, {Elitzur}, \& {Elvis}}]{2006AJ....132..401B}
{Buchanan}, C.~L., {et~al.} 2006, \aj, 132, 401

\bibitem[{{Davidson} \& {Netzer}(1979)}]{1979RvMP...51..715D}
{Davidson}, K., \& {Netzer}, H. 1979, Reviews of Modern Physics, 51, 715

\bibitem[{{Dullemond} \& {van Bemmel}(2005)}]{2005A&A...436...47D}
{Dullemond}, C.~P., \& {van Bemmel}, I.~M. 2005, \aap, 436, 47

\bibitem[{{Edelson} \& {Malkan}(1986)}]{1986ApJ...308...59E}
{Edelson}, R.~A., \& {Malkan}, M.~A. 1986, \apj, 308, 59

\bibitem[{{Elitzur} \& {Shlosman}(2006)}]{2006ApJ...648L.101E}
{Elitzur}, M., \& {Shlosman}, I. 2006, \apjl, 648, L101

\bibitem[{{Fazio} {et~al.}(2004){Fazio}, {Hora}, {Allen}, {Ashby}, {Barmby},  {Deutsch}, {Huang}, {Kleiner}, {Marengo}, {Megeath}, {Melnick}, {Pahre},  {Patten}, {Polizotti}, {Smith}, {Taylor}, {Wang}, {Willner}, {Hoffmann},  {Pipher}, {Forrest}, {McMurty}, {McCreight}, {McKelvey}, {McMurray}, {Koch},  {Moseley}, {Arendt}, {Mentzell}, {Marx}, {Losch}, {Mayman}, {Eichhorn},  {Krebs}, {Jhabvala}, {Gezari}, {Fixsen}, {Flores}, {Shakoorzadeh}, {Jungo},  {Hakun}, {Workman}, {Karpati}, {Kichak}, {Whitley}, {Mann}, {Tollestrup},  {Eisenhardt}, {Stern}, {Gorjian}, {Bhattacharya}, {Carey}, {Nelson},  {Glaccum}, {Lacy}, {Lowrance}, {Laine}, {Reach}, {Stauffer}, {Surace},  {Wilson}, {Wright}, {Hoffman}, {Domingo}, \& {Cohen}}]{2004ApJS..154...10F}
{Fazio}, G.~G., {et~al.} 2004, \apjs, 154, 10

\bibitem[{{Fritz} {et~al.}(2006){Fritz}, {Franceschini}, \&  {Hatziminaoglou}}]{2006MNRAS.366..767F}
{Fritz}, J., {Franceschini}, A., \& {Hatziminaoglou}, E. 2006, \mnras, 366, 767

\bibitem[{{Gallagher} {et~al.}(2007){Gallagher}, {Richards}, {Lacy}, {Hines},  {Elitzur}, \& {Storrie-Lombardi}}]{2007ApJ...661...30G}
{Gallagher}, S.~C., {et~al.} 2007, \apj, 661, 30

\bibitem[{{Glikman} {et~al.}(2006){Glikman}, {Helfand}, \&  {White}}]{2006ApJ...640..579G}
{Glikman}, E., {Helfand}, D.~J., \& {White}, R.~L. 2006, \apj, 640, 579

\bibitem[{{Hao} {et~al.}(2005){Hao}, {Spoon}, {Sloan}, {Marshall}, {Armus},  {Tielens}, {Sargent}, {van Bemmel}, {Charmandaris}, {Weedman}, \&  {Houck}}]{2005ApJ...625L..75H}
{Hao}, L., {et~al.} 2005, \apjl, 625, L75

\bibitem[{{Hatziminaoglou} {et~al.}(2010){Hatziminaoglou}, {Omont}, {Stevens},  {Amblard}, {Arumugam}, {Auld}, {Aussel}, {Babbedge}, {Blain}, {Bock},  {Boselli}, {Buat}, {Burgarella}, {Castro-Rodr{\'{\i}}guez}, {Cava},  {Chanial}, {Clements}, {Conley}, {Conversi}, {Cooray}, {Dowell}, {Dwek},  {Dye}, {Eales}, {Elbaz}, {Farrah}, {Fox}, {Franceschini}, {Gear}, {Glenn},  {Gonz{\'a}lez Solares}, {Griffin}, {Halpern}, {Ibar}, {Isaak}, {Ivison},  {Lagache}, {Levenson}, {Lu}, {Madden}, {Maffei}, {Mainetti}, {Marchetti},  {Mortier}, {Nguyen}, {O'Halloran}, {Oliver}, {Page}, {Panuzzo},  {Papageorgiou}, {Pearson}, {P{\'e}rez-Fournon}, {Pohlen}, {Rawlings},  {Rigopoulou}, {Rizzo}, {Roseboom}, {Rowan-Robinson}, {Sanchez Portal},  {Schulz}, {Scott}, {Seymour}, {Shupe}, {Smith}, {Symeonidis}, {Trichas},  {Tugwell}, {Vaccari}, {Valtchanov}, {Vigroux}, {Wang}, {Ward}, {Wright},  {Xu}, \& {Zemcov}}]{2010A&A...518L..33H}
{Hatziminaoglou}, E., {et~al.} 2010, \aap, 518,  L33+

\bibitem[{{Hewett} \& {Wild}(2010)}]{2010MNRAS.405.2302H}
{Hewett}, P.~C., \& {Wild}, V. 2010, \mnras, 405, 2302

\bibitem[{{Higdon} {et~al.}(2004){Higdon}, {Devost}, {Higdon}, {Brandl},  {Houck}, {Hall}, {Barry}, {Charmandaris}, {Smith}, {Sloan}, \&  {Green}}]{2004PASP..116..975H}
{Higdon}, S.~J.~U., {et~al.} 2004, \pasp, 116, 975

\bibitem[{{H{\"o}nig} {et~al.}(2006){H{\"o}nig}, {Beckert}, {Ohnaka}, \&  {Weigelt}}]{2006A&A...452..459H}
{H{\"o}nig}, S.~F., {et~al.} 2006, \aap,  452, 459

\bibitem[{{Ivezi{\' c}} {et~al.}(1999){Ivezi{\' c}}, {Nenkova}, \&  {Elitzur}}]{1999astro.ph.10475I}
{Ivezi{\' c}}, {\v Z}., {Nenkova}, M., \& {Elitzur}, M. 1999, ArXiv  Astrophysics e-prints

\bibitem[{{Jaffe} {et~al.}(2004){Jaffe}, {Meisenheimer}, {R{\"o}ttgering},  {Leinert}, {Richichi}, {Chesneau}, {Fraix-Burnet}, {Glazenborg-Kluttig},  {Granato}, {Graser}, {Heijligers}, {K{\"o}hler}, {Malbet}, {Miley},  {Paresce}, {Pel}, {Perrin}, {Przygodda}, {Schoeller}, {Sol}, {Waters},  {Weigelt}, {Woillez}, \& {de Zeeuw}}]{2004Natur.429...47J}
{Jaffe}, W., {et~al.} 2004, \nat, 429, 47

\bibitem[{{Konigl} \& {Kartje}(1994)}]{1994ApJ...434..446K}
{Konigl}, A., \& {Kartje}, J.~F. 1994, \apj, 434, 446

\bibitem[{{Krolik} \& {Begelman}(1988)}]{1988ApJ...329..702K}
{Krolik}, J.~H., \& {Begelman}, M.~C. 1988, \apj, 329, 702

\bibitem[{{Laor} \& {Draine}(1993)}]{1993ApJ...402..441L}
{Laor}, A., \& {Draine}, B.~T. 1993, \apj, 402, 441

\bibitem[{{Levenson} {et~al.}(2007){Levenson}, {Sirocky}, {Hao}, {Spoon},  {Marshall}, {Elitzur}, \& {Houck}}]{2007ApJ...654L..45L}
{Levenson}, N.~A., {et~al.} 2007, \apjl, 654, L45

\bibitem[{{Little-Marenin} \& {Little}(1988)}]{1988ApJ...333..305L}
{Little-Marenin}, I.~R., \& {Little}, S.~J. 1988, \apj, 333, 305

\bibitem[{{Lonsdale} {et~al.}(2003){Lonsdale}, {Smith}, {Rowan-Robinson},  {Surace}, {Shupe}, {Xu}, {Oliver}, {Padgett}, {Fang}, {Conrow},  {Franceschini}, {Gautier}, {Griffin}, {Hacking}, {Masci}, {Morrison},  {O'Linger}, {Owen}, {P{\'e}rez-Fournon}, {Pierre}, {Puetter}, {Stacey},  {Castro}, {Polletta}, {Farrah}, {Jarrett}, {Frayer}, {Siana}, {Babbedge},  {Dye}, {Fox}, {Gonzalez-Solares}, {Salaman}, {Berta}, {Condon}, {Dole}, \&  {Serjeant}}]{2003PASP..115..897L}
{Lonsdale}, C.~J., {et~al.} 2003, \pasp, 115, 897

\bibitem[{{Maiolino} {et~al.}(2001){Maiolino}, {Marconi}, \&  {Oliva}}]{2001A&A...365...37M}
{Maiolino}, R., {Marconi}, A., \& {Oliva}, E. 2001, \aap, 365, 37

\bibitem[{{Markwick-Kemper} {et~al.}(2007){Markwick-Kemper}, {Gallagher},  {Hines}, \& {Bouwman}}]{2007ApJ...668L.107M}
{Markwick-Kemper}, F., {et~al.} 2007, \apjl, 668, L107

\bibitem[{{Mason} {et~al.}(2006){Mason}, {Geballe}, {Packham}, {Levenson},  {Elitzur}, {Fisher}, \& {Perlman}}]{2006ApJ...640..612M}
{Mason}, R.~E., {et~al.} 2006, \apj, 640, 612

\bibitem[{{Mor} {et~al.}(2009){Mor}, {Netzer}, \&  {Elitzur}}]{2009ApJ...705..298M}
{Mor}, R., {Netzer}, H., \& {Elitzur}, M. 2009, \apj, 705, 298

\bibitem[{{Murayama} {et~al.}(2000){Murayama}, {Mouri}, \&  {Taniguchi}}]{2000ApJ...528..179M}
{Murayama}, T., {Mouri}, H., \& {Taniguchi}, Y. 2000, \apj, 528, 179

\bibitem[{{Murray} \& {Chiang}(1995)}]{1995ApJ...454L.105M}
{Murray}, N., \& {Chiang}, J. 1995, \apjl, 454, L105+

\bibitem[{{Nenkova} {et~al.}(2002){Nenkova}, {Ivezi{\'c}}, \&  {Elitzur}}]{2002ApJ...570L...9N}
{Nenkova}, M., {Ivezi{\'c}}, {\v Z}., \& {Elitzur}, M. 2002, \apjl, 570, L9

\bibitem[{{Nenkova} {et~al.}(2008a){Nenkova}, {Sirocky},  {Ivezi{\'c}}, \& {Elitzur}}]{2008ApJ...685..147N}
{Nenkova}, M., {et~al.} 2008a, \apj, 685, 147

\bibitem[{{Nenkova} {et~al.}(2008b){Nenkova}, {Sirocky},  {Nikutta}, {Ivezi{\'c}}, \& {Elitzur}}]{2008ApJ...685..160N}
{Nenkova}, M., {et~al.} 2008b, \apj, 685, 160

\bibitem[{{Netzer} {et~al.}(2007){Netzer}, {Lutz}, {Schweitzer}, {Contursi},  {Sturm}, {Tacconi}, {Veilleux}, {Kim}, {Rupke}, {Baker}, {Dasyra},  {Mazzarella}, \& {Lord}}]{2007ApJ...666..806N}
{Netzer}, H., {et~al.} 2007, \apj, 666, 806

\bibitem[{{Neugebauer} {et~al.}(1979){Neugebauer}, {Oke}, {Becklin}, \&  {Matthews}}]{1979ApJ...230...79N}
{Neugebauer}, G., {et~al.} 1979, \apj,  230, 79

\bibitem[{{Nikutta} {et~al.}(2009){Nikutta}, {Elitzur}, \&  {Lacy}}]{2009ApJ...707.1550N}
{Nikutta}, R., {Elitzur}, M., \& {Lacy}, M. 2009, \apj, 707, 1550

\bibitem[{{Pier} \& {Krolik}(1992)}]{1992ApJ...401...99P}
{Pier}, E.~A., \& {Krolik}, J.~H. 1992, \apj, 401, 99

\bibitem[{{Pier} \& {Krolik}(1993)}]{1993ApJ...418..673P}
---. 1993, \apj, 418, 673

\bibitem[{{Polletta} {et~al.}(2008){Polletta}, {Weedman}, {H{\"o}nig},  {Lonsdale}, {Smith}, \& {Houck}}]{2008ApJ...675..960P}
{Polletta}, M., {et~al.} 2008, \apj, 675, 960

\bibitem[{{Proga} {et~al.}(2000){Proga}, {Stone}, \&  {Kallman}}]{2000ApJ...543..686P}
{Proga}, D., {Stone}, J.~M., \& {Kallman}, T.~R. 2000, \apj, 543, 686

\bibitem[{{Rees} {et~al.}(1969){Rees}, {Silk}, {Werner}, \&  {Wickramasinghe}}]{1969Natur.223..788R}
{Rees}, M.~J., {et~al.} 1969,  \nat, 223, 788

\bibitem[{{Richards} {et~al.}(2003){Richards}, {Hall}, {Vanden Berk},  {Strauss}, {Schneider}, {Weinstein}, {Reichard}, {York}, {Knapp}, {Fan},  {Ivezi{\'c}}, {Brinkmann}, {Budav{\'a}ri}, {Csabai}, \&  {Nichol}}]{2003AJ....126.1131R}
{Richards}, G.~T., {et~al.} 2003, \aj, 126, 1131

\bibitem[{{Richards} {et~al.}(2006){Richards}, {Lacy}, {Storrie-Lombardi},  {Hall}, {Gallagher}, {Hines}, {Fan}, {Papovich}, {Vanden Berk}, {Trammell},  {Schneider}, {Vestergaard}, {York}, {Jester}, {Anderson}, {Budav{\'a}ri}, \&  {Szalay}}]{2006ApJS..166..470R}
{Richards}, G.~T., {et~al.} 2006, \apjs, 166, 470

\bibitem[{{Riffel} {et~al.}(2009){Riffel}, {Storchi-Bergmann}, \&  {McGregor}}]{2009ApJ...698.1767R}
{Riffel}, R.~A., {Storchi-Bergmann}, T., \& {McGregor}, P.~J. 2009, \apj, 698,  1767

\bibitem[{{Roche} \& {Aitken}(1984)}]{1984MNRAS.208..481R}
{Roche}, P.~F., \& {Aitken}, D.~K. 1984, \mnras, 208, 481

\bibitem[{{Rodr{\'{\i}}guez-Ardila} \& {Mazzalay}(2006)}]{2006MNRAS.367L..57R}
{Rodr{\'{\i}}guez-Ardila}, A., \& {Mazzalay}, X. 2006, \mnras, 367, L57

\bibitem[{{Rowan-Robinson}(1995)}]{1995MNRAS.272..737R}
{Rowan-Robinson}, M. 1995, \mnras, 272, 737

\bibitem[{{Sanders} {et~al.}(1989){Sanders}, {Phinney}, {Neugebauer}, {Soifer},  \& {Matthews}}]{1989ApJ...347...29S}
{Sanders}, D.~B., {et~al.} 1989, \apj, 347, 29

\bibitem[{{Schartmann} {et~al.}(2005){Schartmann}, {Meisenheimer}, {Camenzind},  {Wolf}, \& {Henning}}]{2005A&A...437..861S}
{Schartmann}, M., {et~al.} 2005, \aap, 437, 861

\bibitem[{{Schartmann} {et~al.}(2008){Schartmann}, {Meisenheimer}, {Camenzind},  {Wolf}, {Tristram}, \& {Henning}}]{2008A&A...482...67S}
{Schartmann}, M., {et~al.} 2008, \aap, 482, 67

\bibitem[{{Schweitzer} {et~al.}(2006){Schweitzer}, {Lutz}, {Sturm}, {Contursi},  {Tacconi}, {Lehnert}, {Dasyra}, {Genzel}, {Veilleux}, {Rupke}, {Kim},  {Baker}, {Netzer}, {Sternberg}, {Mazzarella}, \&  {Lord}}]{2006ApJ...649...79S}
{Schweitzer}, M., {et~al.} 2006, \apj, 649, 79

\bibitem[{{Shi} {et~al.}(2006){Shi}, {Rieke}, {Hines}, {Gorjian}, {Werner},  {Cleary}, {Low}, {Smith}, \& {Bouwman}}]{2006ApJ...653..127S}
{Shi}, Y., {et~al.} 2006, \apj, 653,  127

\bibitem[{{Siebenmorgen} {et~al.}(2005){Siebenmorgen}, {Haas}, {Kruegel}, \&  {Schulz}}]{2005AN....326R.556S}
{Siebenmorgen}, R., {et~al.} 2005,  Astronomische Nachrichten, 326, 556

\bibitem[{{Sturm} {et~al.}(2005){Sturm}, {Schweitzer}, {Lutz}, {Contursi},  {Genzel}, {Lehnert}, {Tacconi}, {Veilleux}, {Rupke}, {Kim}, {Sternberg},  {Maoz}, {Lord}, {Mazzarella}, \& {Sanders}}]{2005ApJ...629L..21S}
{Sturm}, E., {et~al.} 2005, \apjl, 629, L21

\bibitem[{{Tristram} {et~al.}(2007){Tristram}, {Meisenheimer}, {Jaffe},  {Schartmann}, {Rix}, {Leinert}, {Morel}, {Wittkowski}, {R{\"o}ttgering},  {Perrin}, {Lopez}, {Raban}, {Cotton}, {Graser}, {Paresce}, \&  {Henning}}]{2007A&A...474..837T}
{Tristram}, K.~R.~W., {et~al.} 2007, \aap, 474, 837

\bibitem[{{Urry} \& {Padovani}(1995)}]{1995PASP..107..803U}
{Urry}, C.~M., \& {Padovani}, P. 1995, \pasp, 107, 803

\bibitem[{{Weedman} {et~al.}(2005){Weedman}, {Hao}, {Higdon}, {Devost}, {Wu},  {Charmandaris}, {Brandl}, {Bass}, \& {Houck}}]{2005ApJ...633..706W}
{Weedman}, D.~W., {et~al.} 2005, \apj,  633, 706

\bibitem[{{Whittet}(2003)}]{2003dge..conf.....W}
{Whittet}, D.~C.~B., ed. 2003, {Dust in the galactic environment}

\bibitem[{{York} {et~al.}(2000){York}, {Adelman}, {Anderson}, {Anderson},  {Annis}, {Bahcall}, {Bakken}, {Barkhouser}, {Bastian}, {Berman}, {Boroski},  {Bracker}, {Briegel}, {Briggs}, {Brinkmann}, {Brunner}, {Burles}, {Carey},  {Carr}, {Castander}, {Chen}, {Colestock}, {Connolly}, {Crocker}, {Csabai},  {Czarapata}, {Davis}, {Doi}, {Dombeck}, {Eisenstein}, {Ellman}, {Elms},  {Evans}, {Fan}, {Federwitz}, {Fiscelli}, {Friedman}, {Frieman}, {Fukugita},  {Gillespie}, {Gunn}, {Gurbani}, {de Haas}, {Haldeman}, {Harris}, {Hayes},  {Heckman}, {Hennessy}, {Hindsley}, {Holm}, {Holmgren}, {Huang}, {Hull},  {Husby}, {Ichikawa}, {Ichikawa}, {Ivezi{\'c}}, {Kent}, {Kim}, {Kinney},  {Klaene}, {Kleinman}, {Kleinman}, {Knapp}, {Korienek}, {Kron}, {Kunszt},  {Lamb}, {Lee}, {Leger}, {Limmongkol}, {Lindenmeyer}, {Long}, {Loomis},  {Loveday}, {Lucinio}, {Lupton}, {MacKinnon}, {Mannery}, {Mantsch}, {Margon},  {McGehee}, {McKay}, {Meiksin}, {Merelli}, {Monet}, {Munn}, {Narayanan},  {Nash}, {Neilsen}, {Neswold}, {Newberg}, {Nichol}, {Nicinski}, {Nonino},  {Okada}, {Okamura}, {Ostriker}, {Owen}, {Pauls}, {Peoples}, {Peterson},  {Petravick}, {Pier}, {Pope}, {Pordes}, {Prosapio}, {Rechenmacher}, {Quinn},  {Richards}, {Richmond}, {Rivetta}, {Rockosi}, {Ruthmansdorfer}, {Sandford},  {Schlegel}, {Schneider}, {Sekiguchi}, {Sergey}, {Shimasaku}, {Siegmund},  {Smee}, {Smith}, {Snedden}, {Stone}, {Stoughton}, {Strauss}, {Stubbs},  {SubbaRao}, {Szalay}, {Szapudi}, {Szokoly}, {Thakar}, {Tremonti}, {Tucker},  {Uomoto}, {Vanden Berk}, {Vogeley}, {Waddell}, {Wang}, {Watanabe},  {Weinberg}, {Yanny}, \& {Yasuda}}]{2000AJ....120.1579Y}
{York}, D.~G., {et~al.} 2000, \aj, 120, 1579

\end{thebibliography}
  \clearpage


  \begin{deluxetable}{lccrrrrrrrrr}
    \tabletypesize{\tiny}
    \tablewidth{0pt}
    \tablecolumns{12}
    \tablecaption{\textit{Spitzer}/IRS Low Resolution Observation Summary}
    \tablehead{
      \colhead{SDSS} &
      \colhead{\textit{Spitzer}} &
      \colhead{\textit{Spitzer}} &
      \multicolumn{8}{c}{Exposure Time$^a$ (sec.)} &
      \colhead{Pipeline} \\
      \colhead{ID} &
      \colhead{PID} &
      \colhead{AORKEY} &
      \colhead{\#} &
      \colhead{SL2} &
      \colhead{\#} &
      \colhead{SL1} &
      \colhead{\#} &
      \colhead{LL2} &
      \colhead{\#} &
      \colhead{LL1} &
      \colhead{version} \\
      \colhead{(1)} &
      \colhead{(2)} &
      \colhead{(3)} &
      \multicolumn{2}{c}{(4)} &
      \multicolumn{2}{c}{(5)} &
      \multicolumn{2}{c}{(6)} &
      \multicolumn{2}{c}{(7)} &
      \colhead{(8)}
    }
    \startdata
    095047.47+480047.3 & 50328 & 2597 7600 & 3&  60.95 & 5& 14.68 & 2& 121.9 &  4&  31.46 & S18.7.0 \\
    100401.27+423123.0 & 50328 & 2597 6832 & 3&  60.95 & 5& 14.68 & 2& 121.9 &  2& 121.90 & S18.7.0 \\
    103931.14+581709.4 & 50087 & 2538 8544 & 1& 241.83 & 2& 60.95 & 3& 121.9 & 10& 121.90 & S18.7.0 \\
    104114.48+575023.9 & 50087 & 2538 9056 & 1& 241.83 & 2& 60.95 & 4& 121.9 & 10& 121.90 & S18.7.0 \\
    104155.16+571603.0 & 50087 & 2538 8800 & 3&  60.95 & 1& 60.95 & 1& 121.9 &  4& 121.90 & S18.7.0 \\
    104355.49+562757.1 & 50087 & 2538 9312 & 1& 241.83 & 1& 60.95 & 1& 121.9 &  4& 121.90 & S18.7.0 \\
    105001.04+591111.9 & 50087 & 2538 9568 & 1& 241.83 & 2& 60.95 & 2& 121.9 &  9& 121.90 & S18.7.0 \\
    105153.77+565005.7 & 50087 & 2538 8032 & 1& 241.83 & 2& 60.95 & 2& 121.9 &  9& 121.90 & S18.7.0 \\
    105447.28+581909.5 & 50087 & 2538 7264 & 1& 241.83 & 1& 60.95 & 1& 121.9 &  4& 121.90 & S18.7.0 \\
    105951.05+090905.7 & 50328 & 2597 7856 & 8&  14.68 & 3& 14.68 & 2& 121.9 &  4&  31.46 & S18.7.0 \\
    132120.48+574259.4 & 50328 & 2597 7344 & 3&  60.95 & 5& 14.68 & 2& 121.9 &  2& 121.90 & S18.7.0 \\
    142730.19+324106.4 & 50087 & 2538 9824 & 1& 241.83 & 2& 60.95 & 2& 121.9 &  8& 121.90 & S18.7.0 \\
    142954.70+330134.7 & 50087 & 2539 0080 & 3&  60.95 & 1& 60.95 & 1& 121.9 &  4& 121.90 & S18.7.0 \\
    143102.94+323927.8 & 50087 & 2539 0336 & 1& 241.83 & 1& 60.95 & 2& 121.9 &  7& 121.90 & S18.7.0 \\
    143605.07+334242.6 & 50087 & 2539 0592 & 1& 241.83 & 1& 60.95 & 2& 121.9 &  7& 121.90 & S18.7.0 \\
    151307.75+605956.9 & 50328 & 2597 7088 & 2&  60.95 & 5& 14.68 & 2& 121.9 &  2& 121.90 & S18.7.0 \\
    160004.33+550429.9 & 50087 & 2538 8288 & 2& 241.83 & 2& 60.95 & 2& 121.9 &  9& 121.90 & S18.7.0 \\
    160950.72+532909.5 & 50087 & 2539 0848 & 1& 241.83 & 1& 60.95 & 1& 121.9 &  4& 121.90 & S18.7.0 \\
    161007.11+535814.0 &  3640 & 1134 6688 & 2&  60.95 & 2& 60.95 & 2& 121.9 &  2& 121.90 & S18.7.0 \\
    161238.27+532255.0 & 50087 & 2538 7776 & 1& 241.83 & 1& 60.95 & 1& 121.9 &  6& 121.90 & S18.7.0 \\
    163021.65+411147.1 & 50087 & 2538 7520 & 3&  60.95 & 1& 60.95 & 1& 121.9 &  6& 121.90 & S18.7.0 \\
    163425.11+404152.4 &  3640 & 1134 3104 & 2&  60.95 & 2& 60.95 & 2& 121.9 &  2& 121.90 & S18.7.0 \\
    163952.85+410344.8 &  3640 & 1134 5408 & 2&  60.95 & 2& 60.95 & 2& 121.9 &  2& 121.90 & S18.7.0 \\
    164016.08+412101.2 &  3640 & 1134 5920 & 2&  60.95 & 2& 60.95 & 2& 121.9 &  2& 121.90 & S18.7.0 \\
    172522.06+595251.0 & 50087 & 2539 1104 & 2& 241.83 & 2& 60.95 & 2& 121.9 &  9& 121.90 & S18.7.0 \\
    \enddata
    \tablecomments{$a$: The numbers in columns titled ``\#'' give the number of
      spectral images contributing to each observation of a given order and
      nod-position. The exposure times for individual exposures of a nod
      position are given.}
    \label{tab:sample}
  \end{deluxetable}

  \begin{deluxetable}{lccccccccccccc}
    \tabletypesize{\tiny}
    \tablewidth{0pt}
    \tablecolumns{14}
    \tablecaption{SDSS Photometric Measurements.}
    \tablehead{
      \colhead{SDSS ID} &
      \colhead{Redshift} &
      \multicolumn{10}{c}{SDSS} &
      \colhead{$M_{i}$} &
      \colhead{$\Delta(g - i)$} \\
      \colhead{} &
      \colhead{} &
      \multicolumn{2}{c}{$u$} &
      \multicolumn{2}{c}{$g$} &
      \multicolumn{2}{c}{$r$} &
      \multicolumn{2}{c}{$i$} &
      \multicolumn{2}{c}{$z$} &
      \colhead{} &
      \colhead{} \\
      \colhead{} &
      \colhead{} &
      \colhead{Mag.} &
      \colhead{Error} &
      \colhead{Mag.} &
      \colhead{Error} &
      \colhead{Mag.} &
      \colhead{Error} &
      \colhead{Mag.} &
      \colhead{Error} &
      \colhead{Mag.} &
      \colhead{Error} &
      \colhead{} &
      \colhead{}
    } \startdata
    095047.47+480047.3 & 1.743280 & 17.247 & 0.019 & 17.183 & 0.021 & 17.118 & 0.016 & 16.837 & 0.016 & 16.764 & 0.014 & -28.222 &  0.117  \\
    100401.27+423123.0 & 1.653350 & 17.06  & 0.021 & 16.857 & 0.023 & 16.883 & 0.017 & 16.764 & 0.014 & 16.772 & 0.027 & -28.187 & -0.161  \\
    103931.14+581709.4 & 1.829790 & 18.439 & 0.033 & 18.452 & 0.048 & 18.442 & 0.023 & 18.19  & 0.014 & 18.228 & 0.025 & -26.98  &  0.028  \\
    104114.48+575023.9 & 1.902640 & 19.012 & 0.026 & 18.875 & 0.016 & 18.919 & 0.027 & 18.737 & 0.018 & 18.759 & 0.037 & -26.522 & -0.072  \\
    104155.16+571603.0 & 1.720740 & 17.944 & 0.025 & 17.893 & 0.01  & 17.921 & 0.014 & 17.735 & 0.017 & 17.721 & 0.02  & -27.288 & -0.08   \\
    104355.49+562757.1 & 1.947830 & 17.767 & 0.022 & 17.667 & 0.017 & 17.479 & 0.023 & 17.188 & 0.01  & 17.075 & 0.024 & -28.124 &  0.296  \\
    105001.04+591111.9 & 2.167560 & 19.798 & 0.045 & 19.432 & 0.029 & 19.209 & 0.02  & 19.087 & 0.024 & 18.868 & 0.038 & -26.474 &  0.229  \\
    105153.77+565005.7 & 1.975930 & 18.721 & 0.019 & 18.78  & 0.015 & 18.706 & 0.015 & 18.549 & 0.019 & 18.371 & 0.027 & -26.803 &  0.056  \\
    105447.28+581909.5 & 1.653240 & 18.28  & 0.017 & 18.058 & 0.03  & 18.008 & 0.014 & 17.763 & 0.011 & 17.82  & 0.036 & -27.168 &  0.058  \\
    105951.05+090905.7 & 1.688240 & 17.214 & 0.022 & 17.243 & 0.032 & 17.052 & 0.017 & 16.771 & 0.017 & 16.8   & 0.036 & -28.25  &  0.19   \\
    132120.48+574259.4 & 1.773950 & 17.205 & 0.036 & 17.139 & 0.022 & 17.069 & 0.013 & 16.842 & 0.012 & 16.85  & 0.026 & -28.271 &  0.06   \\
    142730.19+324106.4 & 1.775950 & 19.425 & 0.036 & 19.186 & 0.024 & 19.015 & 0.016 & 18.886 & 0.015 & 18.835 & 0.041 & -26.225 &  0.061  \\
    142954.70+330134.7 & 2.075990 & 18.467 & 0.02  & 18.352 & 0.024 & 18.247 & 0.013 & 18.098 & 0.015 & 17.916 & 0.032 & -27.362 &  0.115  \\
    143102.94+323927.8 & 1.643710 & 18.603 & 0.018 & 18.436 & 0.015 & 18.298 & 0.019 & 18.111 & 0.014 & 18.119 & 0.027 & -26.812 &  0.067  \\
    143605.07+334242.6 & 1.986070 & 18.609 & 0.035 & 18.595 & 0.023 & 18.511 & 0.012 & 18.334 & 0.021 & 18.19  & 0.029 & -27.028 &  0.089  \\
    151307.75+605956.9 & 1.822110 & 17.022 & 0.016 & 16.945 & 0.021 & 16.892 & 0.015 & 16.705 & 0.015 & 16.69  & 0.023 & -28.47  & -0.02   \\
    160004.33+550429.9 & 1.982860 & 18.962 & 0.03  & 18.858 & 0.016 & 18.823 & 0.019 & 18.792 & 0.019 & 18.703 & 0.033 & -26.568 & -0.107  \\
    160950.72+532909.5 & 1.716120 & 18.161 & 0.026 & 18.046 & 0.021 & 18.043 & 0.024 & 17.869 & 0.022 & 17.796 & 0.024 & -27.158 & -0.07   \\
    161007.11+535814.0 & 2.030270 & 19.009 & 0.033 & 18.938 & 0.018 & 18.858 & 0.019 & 18.785 & 0.022 & 18.563 & 0.034 & -26.631 & -0.018  \\
    161238.27+532255.0 & 2.139160 & 17.95  & 0.031 & 17.839 & 0.022 & 17.826 & 0.017 & 17.728 & 0.018 & 17.478 & 0.023 & -27.811 & -0.01   \\
    163021.65+411147.1 & 1.646520 & 18.435 & 0.018 & 18.262 & 0.011 & 18.259 & 0.017 & 18.072 & 0.014 & 18.149 & 0.029 & -26.861 & -0.069  \\
    163425.11+404152.4 & 1.692170 & 18.531 & 0.023 & 18.409 & 0.018 & 18.372 & 0.015 & 18.136 & 0.013 & 18.169 & 0.033 & -26.853 &  0.04   \\
    163952.85+410344.8 & 1.602630 & 18.8   & 0.027 & 18.638 & 0.022 & 18.589 & 0.018 & 18.35  & 0.013 & 18.452 & 0.031 & -26.512 &  0.018  \\
    164016.08+412101.2 & 1.761550 & 18.878 & 0.022 & 18.596 & 0.012 & 18.438 & 0.016 & 18.06  & 0.016 & 17.98  & 0.022 & -27.025 &  0.305  \\
    172522.06+595251.0 & 1.872150 & 19.347 & 0.035 & 19.164 & 0.028 & 18.902 & 0.017 & 18.774 & 0.025 & 18.744 & 0.046 & -26.497 &  0.093  \\
    \enddata
    \tablecomments{SDSS measurements are taken from the SDSS DR7 database. The
      photometry is corrected for Galactic extinction. The redshifts are taken
      from the work of \citet{2010MNRAS.405.2302H}}
    \label{tab:continua}
  \end{deluxetable}

  \begin{deluxetable}{lp{0.2cm}p{0.2cm}p{0.2cm}p{0.2cm}p{0.2cm}p{0.2cm}p{0.2cm}p{0.2cm}p{0.2cm}p{0.2cm}p{0.2cm}p{0.2cm}p{0.2cm}p{0.2cm}p{0.2cm}p{0.2cm}}
    \tabletypesize{\tiny}
    \tablewidth{0pt}
    \tablecolumns{17}
    \tablecaption{2MASS, IRAC and MIPS Photometric Measurements.}
    \tablehead{
      \colhead{SDSS ID} &
      \multicolumn{6}{c}{2MASS} &
      \multicolumn{8}{c}{IRAC ($\mu Jy$)} &
      \multicolumn{2}{c}{MIPS ($\mu Jy$)} \\
      \colhead{} &
      \multicolumn{2}{c}{$J$} &
      \multicolumn{2}{c}{$H$} &
      \multicolumn{2}{c}{$K_{s}$} &
      \multicolumn{2}{c}{3.6$\mum$} &
      \multicolumn{2}{c}{4.5$\mum$} &
      \multicolumn{2}{c}{5.8$\mum$} &
      \multicolumn{2}{c}{8.0$\mum$} &
      \multicolumn{2}{c}{24$\mum$} \\
      \colhead{} &
      \colhead{Mag} &
      \colhead{Err} &
      \colhead{Mag} &
      \colhead{Err} &
      \colhead{Mag} &
      \colhead{Err} &
      \colhead{Flux} &
      \colhead{Err} &
      \colhead{Flux} &
      \colhead{Err} &
      \colhead{Flux} &
      \colhead{Err} &
      \colhead{Flux} &
      \colhead{Err} &
      \colhead{Flux} &
      \colhead{Err}
    }
    \startdata
    095047.47+480047.3 & 15.870   & 0.084    & 15.127   & 0.095    & 14.847   & 0.117    & $\ldots$ & $\ldots$ & $\ldots$ & $\ldots$ & $\ldots$ & $\ldots$ & $\ldots$ & $\ldots$ & $\ldots$ & $\ldots$ \\
    100401.27+423123.0 & 15.795   & 0.065    & 15.376   & 0.082    & 15.080   & 0.114    & $\ldots$ & $\ldots$ & $\ldots$ & $\ldots$ & $\ldots$ & $\ldots$ & $\ldots$ & $\ldots$ & $\ldots$ & $\ldots$ \\
    103931.14+581709.4 & 17.380   & 0.090    & 17.254   & 0.212    & 16.794   & 0.258    & 259.19   & 2.21     & 387.72   & 2.95     & 699.91   & 9.58     & 1067.13  & 8.27     & 2723.00  & 23.18    \\
    104114.48+575023.9 & 17.836   & 0.159    & 17.439   & 0.218    & $\ldots$ & $\ldots$ & 186.88   & 1.69     & 331.45   & 2.29     & 585.95   & 8.16     & 1029.81  & 6.67     & 2118.87  & 21.61    \\
    104155.16+571603.0 & 16.846   & 0.073    & 16.511   & 0.153    & 15.963   & 0.144    & 430.66   & 2.76     & 674.89   & 3.18     & 1163.16  & 11.43    & 1802.76  & 8.51     & 4895.25  & 24.09    \\
    104355.49+562757.1 & 16.373   & 0.114    & 16.092   & 0.188    & 15.407   & 0.164    & 580.68   & 3.16     & 780.90   & 5.27     & 1422.47  & 12.41    & 2506.98  & 14.06    & 6799.12  & 21.44    \\
    105001.04+591111.9 & $\ldots$ & $\ldots$ & $\ldots$ & $\ldots$ & $\ldots$ & $\ldots$ & 199.25   & 1.85     & 336.71   & 2.65     & 694.99   & 9.04     & 1226.55  & 8.09     & 3128.45  & 24.09    \\
    105153.77+565005.7 & 17.588   & 0.123    & 17.078   & 0.222    & 16.334   & 0.192    & 262.89   & 1.93     & 427.21   & 2.88     & 782.47   & 8.62     & 1314.91  & 7.60     & 3131.75  & 17.76    \\
    105447.28+581909.5 & 16.906   & 0.089    & 16.222   & 0.136    & 15.986   & 0.161    & 561.02   & 2.48     & 982.87   & 4.65     & 1718.43  & 10.07    & 2973.93  & 12.12    & 8461.48  & 18.10    \\
    105951.05+090905.7 & 15.620   & 0.103    & 15.094   & 0.096    & 14.411   & 0.106    & $\ldots$ & $\ldots$ & $\ldots$ & $\ldots$ & $\ldots$ & $\ldots$ & $\ldots$ & $\ldots$ & $\ldots$ & $\ldots$ \\
    132120.48+574259.4 & 16.201   & 0.077    & 15.523   & 0.092    & 15.058   & 0.098    & $\ldots$ & $\ldots$ & $\ldots$ & $\ldots$ & $\ldots$ & $\ldots$ & $\ldots$ & $\ldots$ & $\ldots$ & $\ldots$ \\
    142730.19+324106.4 & 17.798   & 0.340    & $\ldots$ & $\ldots$ & $\ldots$ & $\ldots$ & $\ldots$ & $\ldots$ & $\ldots$ & $\ldots$ & $\ldots$ & $\ldots$ & $\ldots$ & $\ldots$ & $\ldots$ & $\ldots$ \\
    142954.70+330134.7 & 17.355   & 0.302    & 16.586   & 0.289    & 15.783   & 0.229    & $\ldots$ & $\ldots$ & $\ldots$ & $\ldots$ & $\ldots$ & $\ldots$ & $\ldots$ & $\ldots$ & $\ldots$ & $\ldots$ \\
    143102.94+323927.8 & 17.205   & 0.273    & 17.127   & 0.348    & $\ldots$ & $\ldots$ & $\ldots$ & $\ldots$ & $\ldots$ & $\ldots$ & $\ldots$ & $\ldots$ & $\ldots$ & $\ldots$ & $\ldots$ & $\ldots$ \\
    143605.07+334242.6 & 17.455   & 0.321    & $\ldots$ & $\ldots$ & $\ldots$ & $\ldots$ & $\ldots$ & $\ldots$ & $\ldots$ & $\ldots$ & $\ldots$ & $\ldots$ & $\ldots$ & $\ldots$ & $\ldots$ & $\ldots$ \\
    151307.75+605956.9 & 15.952   & 0.081    & 15.681   & 0.134    & 14.949   & 0.141    & $\ldots$ & $\ldots$ & $\ldots$ & $\ldots$ & $\ldots$ & $\ldots$ & $\ldots$ & $\ldots$ & $\ldots$ & $\ldots$ \\
    160004.33+550429.9 & $\ldots$ & $\ldots$ & $\ldots$ & $\ldots$ & $\ldots$ & $\ldots$ & 195.87   & 1.06     & 313.01   & 1.74     & 579.79   & 5.44     & 1059.13  & 5.50     & 3407.93  & 21.49    \\
    160950.72+532909.5 & 16.924   & 0.247    & $\ldots$ & $\ldots$ & $\ldots$ & $\ldots$ & 572.26   & 2.14     & 965.53   & 3.09     & 1588.99  & 8.58     & 3014.96  & 8.33     & 5991.64  & 21.93    \\
    161007.11+535814.0 & 17.036   & 0.281    & $\ldots$ & $\ldots$ & $\ldots$ & $\ldots$ & 182.13   & 1.87     & 261.76   & 2.19     & 525.77   & 8.52     & 958.75   & 6.19     & 3564.07  & 20.16    \\
    161238.27+532255.0 & 16.698   & 0.151    & 16.146   & 0.225    & 15.542   & 0.228    & 375.80   & 1.73     & 476.13   & 2.27     & 811.68   & 6.76     & 1334.74  & 6.56     & 3705.08  & 20.51    \\
    163021.65+411147.1 & 17.179   & 0.275    & 16.159   & 0.229    & 15.955   & 0.248    & 285.71   & 1.69     & 526.08   & 3.12     & 897.45   & 7.52     & 1568.20  & 8.17     & 3594.86  & 20.58    \\
    163425.11+404152.4 & 17.032   & 0.240    & 16.119   & 0.215    & $\ldots$ & $\ldots$ & 352.76   & 2.05     & 637.06   & 3.12     & 1075.56  & 8.87     & 1915.82  & 7.48     & 4370.81  & 20.44    \\
    163952.85+410344.8 & $\ldots$ & $\ldots$ & $\ldots$ & $\ldots$ & $\ldots$ & $\ldots$ & 253.40   & 1.81     & 406.55   & 2.34     & 617.09   & 8.52     & 983.45   & 5.77     & 2126.06  & 21.26    \\
    164016.08+412101.2 & 16.767   & 0.231    & $\ldots$ & $\ldots$ & $\ldots$ & $\ldots$ & 329.74   & 2.05     & 452.99   & 3.02     & 754.74   & 8.64     & 1281.08  & 7.70     & 3223.61  & 19.91    \\
    172522.06+595251.0 & $\ldots$ & $\ldots$ & $\ldots$ & $\ldots$ & $\ldots$ & $\ldots$ & 195.2    & 20.3     & 356.4    & 36.7     & 597.8    & 63.3     & 1058.4   & 108.1    & 2240.00  & 60.00    \\
    \enddata
    \tablecomments{The 2MASS measurements are from the 2MASS database. The IRAC
      and MIPS fluxes are from the SWIRE catalogs.}
    \label{tab:continua2}
  \end{deluxetable}

  \begin{deluxetable}{lcccccccc}
    \tabletypesize{\tiny}
    \tablewidth{0pt}
    \tablecolumns{9}
    \tablecaption{IRS Photometric Measurements.}
    \tablehead{
      \colhead{SDSS ID} &
      \multicolumn{2}{c}{$3.0\,\mum$} &
      \multicolumn{2}{c}{$5.0\,\mum$} &
      \multicolumn{2}{c}{$8.0\,\mum$} &
      \multicolumn{2}{c}{$10.0\,\mum$} \\
      \colhead{} &
      \colhead{Flux} &
      \colhead{Error} &
      \colhead{Flux} &
      \colhead{Error} &
      \colhead{Flux} &
      \colhead{Error} &
      \colhead{Flux} &
      \colhead{Error}
    }
    \startdata
    095047.47+480047.3 & 2328.64  & 272.65   & 3441.50  & 602.14   & 4776.98  & 365.58   & 6611.23  & 499.03   \\
    100401.27+423123.0 & 5470.28  & 643.56   & 8439.88  & 398.89   & 12556.96 & 787.38   & 17481.36 & 1061.88  \\
    103931.14+581709.4 & 1003.13  & 206.63   & 1612.28  & 166.41   & 2477.67  & 231.32   & 3363.35  & 250.54   \\
    104114.48+575023.9 & 972.94   & 126.76   & 1429.34  & 114.42   & 1967.45  & 107.94   & 3175.65  & 210.57   \\
    104155.16+571603.0 & 1793.85  & 272.61   & 2905.40  & 329.94   & 4176.98  & 330.18   & 6193.95  & 560.67   \\
    104355.49+562757.1 & 2437.01  & 460.34   & 4582.81  & 357.18   & 6344.89  & 359.40   & 9850.30  & 604.97   \\
    105001.04+591111.9 & 1386.64  & 247.97   & 2355.79  & 276.77   & 3361.15  & 329.66   & 3622.24  & 309.14   \\
    105153.77+565005.7 & 1374.48  & 224.88   & 2178.74  & 166.25   & 3137.47  & 227.87   & 4452.11  & 178.69   \\
    105447.28+581909.5 & 2556.06  & 381.77   & 4411.63  & 354.33   & 7146.42  & 355.39   & 10866.61 & 579.99   \\
    105951.05+090905.7 & 2532.34  & 478.23   & 4875.93  & 370.53   & 7338.59  & 743.63   & 11297.45 & 795.49   \\
    132120.48+574259.4 & 2918.96  & 288.83   & 4256.87  & 339.46   & 4719.35  & 236.63   & 6354.48  & 322.14   \\
    142730.19+324106.4 & 1298.06  & 255.39   & 3156.97  & 475.51   & 7706.25  & 393.72   & 5371.36  & 263.97   \\
    142954.70+330134.7 & 1642.35  & 316.31   & 2815.09  & 409.52   & 4487.82  & 317.82   & 6364.06  & 555.97   \\
    143102.94+323927.8 & 1115.51  & 241.66   & 1956.42  & 374.22   & 2814.35  & 329.59   & 4008.50  & 287.88   \\
    143605.07+334242.6 & 1286.07  & 285.40   & 2106.83  & 214.65   & 2927.30  & 385.56   & 4542.41  & 455.34   \\
    151307.75+605956.9 & 4504.23  & 511.16   & 6181.79  & 287.52   & 8024.15  & 434.28   & 12082.52 & 356.26   \\
    160004.33+550429.9 & 1074.17  & 194.95   & 1668.38  & 206.21   & 3398.58  & 218.03   & 3882.49  & 227.51   \\
    160950.72+532909.5 & 2589.88  & 410.45   & 4199.97  & 420.71   & 5746.11  & 403.21   & 6461.92  & 375.45   \\
    161007.11+535814.0 & 1102.34  & 227.58   & 2112.49  & 356.99   & 3713.69  & 433.79   & 6720.09  & 677.66   \\
    161238.27+532255.0 & 1566.17  & 254.13   & 2806.43  & 314.43   & 3971.20  & 253.56   & 5494.72  & 454.38   \\
    163021.65+411147.1 & $\ldots$ & $\ldots$ & $\ldots$ & $\ldots$ & $\ldots$ & $\ldots$ & $\ldots$ & $\ldots$ \\
    163425.11+404152.4 & 1663.63  & 291.83   & 2804.37  & 414.97   & 3818.12  & 470.48   & 5016.83  & 634.67   \\
    163952.85+410344.8 & 878.12   & 206.47   & 1346.60  & 260.96   & 1898.77  & 381.28   & 2714.11  & 420.19   \\
    164016.08+412101.2 & 1159.42  & 222.58   & 1782.03  & 272.68   & 2857.22  & 455.87   & 3856.00  & 596.21   \\
    172522.06+595251.0 & 999.62   & 181.64   & 1532.41  & 132.03   & 2232.69  & 282.91   & 2253.27  & 178.34   \\
    \enddata
    \tablecomments{Each continuum measurement is the error-weighted average of
      the flux densities within a window of 1 $\mum$ centered on the
      respective wavelength. The continuum measurements are obtained with
      deredshifted spectra and are in units of $\mu$Jy}
    \label{tab:continua3}
  \end{deluxetable}

  \begin{deluxetable}{lrrrrrrrr}
    \tabletypesize{\tiny}
    \tablewidth{0pt}
    \tablecolumns{9}
    \tablecaption{Best-fit \clumpy Torus Parameters without an additional blackbody component.}
    \tablehead{
      \colhead{SDSS ID} &
      \colhead{$q$} &
      \colhead{$N_{0}$} &
      \colhead{$\tau_{V}$} &
      \colhead{$Y$} &
      \colhead{$\sigma$} &
      \colhead{$i$} &
      \colhead{$\chi_{\nu}^{2}$} &
      \colhead{$E_{Min}$} \\
      \colhead{} &
      \colhead{} &
      \colhead{} &
      \colhead{} &
      \colhead{} &
      \colhead{} &
      \colhead{} &
      \colhead{} &
      \colhead{}
    }
    \startdata
    095047.47+480047.3 &  3.0 &  1 &   60.0 &   5 &   15 &  90 &  2.3023 &   0.2086 \\
    100401.27+423123.0 &  3.0 &  1 &   80.0 &   5 &   30 &  60 &  7.5916 &   0.3655 \\
    103931.14+581709.4 &  2.0 &  2 &   40.0 &   5 &   15 &  80 &  1.0076 &   0.1420 \\
    104114.48+575023.9 &  3.0 &  1 &   30.0 &   5 &   30 &  80 &  1.7045 &   0.1829 \\
    104155.16+571603.0 &  1.0 &  1 &   60.0 &   5 &   25 &  80 &  4.4097 &   0.2914 \\
    104355.49+562757.1 &  1.0 &  3 &   20.0 &   5 &   15 &  90 &  7.0194 &   0.3577 \\
    105001.04+591111.9 &  3.0 &  2 &   150.0 &  5 &   35 &  40 &  7.2815 &   0.3710 \\
    105153.77+565005.7 &  2.5 &  1 &   40.0 &   5 &   35 &  70 &  4.6752 &   0.3029 \\
    105447.28+581909.5 &  2.5 &  2 &   80.0 &   10 &  30 &  60 &  2.4313 &   0.2124 \\
    105951.05+090905.7 &  3.0 &  1 &   100.0 &  5 &   25 &  90 &  0.7722 &   0.1220 \\
    132120.48+574259.4 &  3.0 &  1 &   40.0 &   5 &   15 &  90 &  8.0009 &   0.3889 \\
    142730.19+324106.4 &  0.5 &  15 &  10.0 &   5 &   35 &  80 &  2.3793 &   0.2161 \\
    142954.70+330134.7 &  3.0 &  1 &   60.0 &   5 &   35 &  70 &  0.8621 &   0.1301 \\
    143102.94+323927.8 &  2.0 &  4 &   30.0 &   60 &  15 &  80 &  0.8203 &   0.1245 \\
    143605.07+334242.6 &  3.0 &  1 &   80.0 &   10 &  30 &  60 &  1.8428 &   0.1866 \\
    151307.75+605956.9 &  3.0 &  1 &   100.0 &  5 &   20 &  70 &  8.4899 &   0.3899 \\
    160004.33+550429.9 &  3.0 &  4 &   20.0 &   60 &  20 &  80 &  3.3760 &   0.2574 \\
    160950.72+532909.5 &  3.0 &  1 &   100.0 &  5 &   30 &  40 &  4.8608 &   0.3089 \\
    161007.11+535814.0 &  0.0 &  1 &   40.0 &   5 &   60 &  40 &  2.5430 &   0.2234 \\
    161238.27+532255.0 &  2.5 &  2 &   10.0 &   40 &  15 &  0 &  38.6937 &   0.8475 \\
    163021.65+411147.1 &  1.5 &  2 &   10.0 &   10 &  15 &  0 &  90.6623 &   1.9043 \\
    163425.11+404152.4 &  3.0 &  1 &   40.0 &   5 &   30 &  70 &  1.0808 &   0.1429 \\
    163952.85+410344.8 &  2.0 &  1 &   150.0 &  5 &   20 &  20 &  1.0997 &   0.1455 \\
    164016.08+412101.2 &  3.0 &  2 &   60.0 &   5 &   15 &  70 &  5.2334 &   0.3205 \\
    172522.06+595251.0 &  3.0 &  3 &   20.0 &   5 &   15 &  80 &  1.4518 &   0.1688 \\
    \enddata
    \tablecomments{Descriptions of \clumpy torus parameters: $q$: index of the
      radial distribution ($r^{-q}$) of clouds; $N_{0}$: average number of
      clouds along radial equatorial rays; $\tau_{V}$: optical depth through a
      single cloud at optical wavelengths; $Y$: the ratio of outer to inner
      (sublimation) radius of the torus.; $\sigma$: the angular width of the
      torus in degrees; $i$: inclination of line-of-sight of the observer in
      degrees; The $\chi_{\nu}^{2}$ and the $E_{Min}$ provide measures of how
      well the best-fit model fits the observed data. Typically
      $\chi_{\nu}^{2}$ close to 1 and smaller values of $E_{Min}$ indicate a
      better fit.}
    \label{tab:torus_best_fit_without_bb}
  \end{deluxetable}

  \begin{deluxetable}{lrrrrrrrrr}
    \tabletypesize{\tiny}
    \tablewidth{0pt}
    \tablecolumns{10}
    \tablecaption{Best-fit \clumpy Torus Parameters for a model with an
      additional blackbody component.}
    \tablehead{
      \colhead{SDSS ID} &
      \colhead{$q$} &
      \colhead{$N_{0}$} &
      \colhead{$\tau_{V}$} &
      \colhead{$Y$} &
      \colhead{$\sigma$} &
      \colhead{$i$} &
      \colhead{$T_{BB}$} &
      \colhead{$\chi_{\nu}^{2}$} &
      \colhead{$E_{Min}$}\\
      \colhead{} &
      \colhead{} &
      \colhead{} &
      \colhead{} &
      \colhead{} &
      \colhead{} &
      \colhead{} &
      \colhead{} &
      \colhead{} &
      \colhead{}
    }
    \startdata
    095047.47+480047.3 &   3.0 &   1 &    100.0 &   100 &   15 &   90 &   1361.7 &  1.5393 &    0.1684\\
    100401.27+423123.0 &   0.0 &   2 &    30.0 &    5 &     15 &   80 &   1165.3 &  5.3058 &    0.3021\\
    103931.14+581709.4 &   1.0 &   3 &    60.0 &    40 &    15 &   80 &   1160.5 &  0.5226 &    0.1009\\
    104114.48+575023.9 &   1.0 &   2 &    80.0 &    10 &    15 &   70 &   1192.6 &  1.5244 &    0.1707\\
    104155.16+571603.0 &   0.5 &   3 &    20.0 &    80 &    15 &   90 &   1203.6 &  2.5872 &    0.2203\\
    104355.49+562757.1 &   0.0 &   2 &    30.0 &    70 &    15 &   90 &   923.8 &   5.1286 &    0.3021\\
    105001.04+591111.9 &   3.0 &   6 &    80.0 &    5 &     25 &   90 &   1320.2 &  1.6741 &    0.1757\\
    105153.77+565005.7 &   0.5 &   11 &   60.0 &    90 &    20 &   60 &   1262.6 &  4.0436 &    0.2780\\
    105447.28+581909.5 &   1.0 &   2 &    30.0 &    90 &    35 &   90 &   1370.8 &  0.8325 &    0.1228\\
    105951.05+090905.7 &   0.5 &   2 &    150.0 &   5 &     15 &   90 &   821.7 &   0.2711 &    0.0713\\
    132120.48+574259.4 &   1.5 &   10 &   150.0 &   90 &    15 &   80 &   1035.6 &  3.6275 &    0.2586\\
    142730.19+324106.4 &   2.5 &   14 &   30.0 &    90 &    35 &   80 &   884.4 &   0.3900 &    0.0863\\
    142954.70+330134.7 &   1.5 &   4 &    60.0 &    40 &    15 &   70 &   1155.8 &  0.7038 &    0.1160\\
    143102.94+323927.8 &   0.5 &   4 &    30.0 &    50 &    15 &   80 &   1156.6 &  0.1212 &    0.0473\\
    143605.07+334242.6 &   0.0 &   2 &    40.0 &    100 &   25 &   80 &   1104.4 &  1.0538 &    0.1394\\
    151307.75+605956.9 &   0.5 &   1 &    40.0 &    10 &    15 &   90 &   1273.5 &  4.8365 &    0.2909\\
    160004.33+550429.9 &   0.0 &   9 &    100.0 &   90 &    65 &   0 &    1093.7 &  2.7757 &    0.2303\\
    160950.72+532909.5 &   2.0 &   3 &    40.0 &    100 &   20 &   90 &   1357.5 &  0.7781 &    0.1219\\
    161007.11+535814.0 &   0.0 &   2 &    60.0 &    80 &    70 &   40 &   1165.8 &  2.1844 &    0.2043\\
    161238.27+532255.0 &   2.0 &   14 &   150.0 &   100 &   60 &   90 &   1168.2 &  35.3965 &   0.8006\\
    163021.65+411147.1 &   2.0 &   15 &   150.0 &   90 &    65 &   60 &   1192.2 &  80.4726 &   1.5538\\
    163425.11+404152.4 &   2.0 &   3 &    80.0 &    100 &   20 &   80 &   1218.1 &  0.5393 &    0.0997\\
    163952.85+410344.8 &   0.0 &   2 &    40.0 &    20 &    15 &   90 &   1348.2 &  0.8820 &    0.1286\\
    164016.08+412101.2 &   0.0 &   15 &   40.0 &    30 &    15 &   70 &   1394.1 &  5.0215 &    0.3098\\
    172522.06+595251.0 &   3.0 &   4 &    60.0 &    20 &    15 &   90 &   1265.7 &  0.3355 &    0.0801\\
    \enddata
    \tablecomments{This table presents best fit \clumpy parameters with an
      additional blackbody component. For descriptions of \clumpy torus
      parameters, please see notes to
      Table~\ref{tab:torus_best_fit_without_bb}. $T_{BB}$ is the temperature
      of the blackbody component in Kelvins.}
    \label{tab:torus_best_fit_with_bb}
  \end{deluxetable}

  \begin{deluxetable}{lr|rrr|rrr|rrr|rrr}
    \tabletypesize{\tiny}
    \tablewidth{0pt}
    \tablecolumns{14}
    \tablecaption{Statistics on the parameter distributions for \textsc{Clumpy}$+$Blackbody model.}
    \tablehead{
      \colhead{SDSS ID} &
      \colhead{Number} &
      \multicolumn{3}{c}{$q$} &
      \multicolumn{3}{c}{$N_{0}$} &
      \multicolumn{3}{c}{$\tau_{V}$} &
      \multicolumn{3}{c}{$\sigma$} \\
      \colhead{} &
      \colhead{of models} &
      \colhead{Mode} &
      \colhead{90\%} &
      \colhead{$D_{KL}$} &
      \colhead{Mode} &
      \colhead{90\%} &
      \colhead{$D_{KL}$} &
      \colhead{Mode} &
      \colhead{90\%} &
      \colhead{$D_{KL}$} &
      \colhead{Mode} &
      \colhead{90\%} &
      \colhead{$D_{KL}$}
    }
    \startdata
    095047.47+480047.3 & 668    & 0.0 & 0.0 2.5 & 0.26 & 7  & 1 13 & 0.06 & 150.0 & 100.0 150.0 & 0.62 & 15 & 15 35 & 0.32 \\
    100401.27+423123.0 & 883    & 3.0 & 1.5 3.0 & 0.24 & 1  & 1 2  & 0.69 & 80.0  & 30.0 150.0  & 0.19 & 25 & 15 30 & 0.50 \\
    103931.14+581709.4 & 1857   & 0.0 & 0.0 2.5 & 0.05 & 4  & 2 13 & 0.04 & 100.0 & 40.0 150.0  & 0.27 & 15 & 15 40 & 0.29 \\
    104114.48+575023.9 & 7911   & 1.5 & 0.0 2.5 & 0.04 & 1  & 1 10 & 0.18 & 150.0 & 30.0 150.0  & 0.20 & 20 & 15 40 & 0.24 \\
    104155.16+571603.0 & 1392   & 0.0 & 0.0 2.5 & 0.10 & 3  & 2 12 & 0.09 & 150.0 & 30.0 150.0  & 0.16 & 15 & 15 25 & 0.51 \\
    104355.49+562757.1 & 3888   & 0.0 & 0.0 2.0 & 0.18 & 5  & 2 14 & 0.02 & 150.0 & 60.0 150.0  & 0.40 & 20 & 15 50 & 0.14 \\
    105001.04+591111.9 & 27260  & 3.0 & 0.0 3.0 & 0.01 & 4  & 1 13 & 0.02 & 80.0  & 20.0 100.0  & 0.16 & 15 & 15 55 & 0.08 \\
    105153.77+565005.7 & 24033  & 2.0 & 0.0 2.5 & 0.01 & 2  & 1 13 & 0.05 & 150.0 & 30.0 150.0  & 0.16 & 15 & 15 55 & 0.12 \\
    105447.28+581909.5 & 63     & 1.0 & 0.5 1.5 & 0.36 & 2  & 2 7  & 0.39 & 30.0  & 20.0 60.0   & 0.36 & 15 & 15 35 & 0.40 \\
    105951.05+090905.7 & 28     & 3.0 & 0.5 3.0 & 0.21 & 3  & 2 3  & 0.76 & 150.0 & 150.0 150.0 & 0.88 & 15 & 15 15 & 0.90 \\
    132120.48+574259.4 & 6113   & 0.0 & 0.0 1.0 & 0.32 & 9  & 6 15 & 0.10 & 150.0 & 80.0 150.0  & 0.45 & 15 & 15 50 & 0.12 \\
    142730.19+324106.4 & 281    & 2.5 & 2.0 2.5 & 0.52 & 15 & 8 15 & 0.21 & 30.0  & 20.0 30.0   & 0.68 & 35 & 35 40 & 0.68 \\
    142954.70+330134.7 & 3537   & 3.0 & 0.0 3.0 & 0.00 & 2  & 1 11 & 0.13 & 150.0 & 40.0 150.0  & 0.25 & 15 & 15 35 & 0.32 \\
    143102.94+323927.8 & 21     & 0.5 & 0.0 1.0 & 0.56 & 4  & 3 4  & 0.69 & 30.0  & 20.0 40.0   & 0.48 & 15 & 15 15 & 0.92 \\
    143605.07+334242.6 & 3765   & 0.0 & 0.0 2.0 & 0.12 & 5  & 2 13 & 0.05 & 150.0 & 60.0 150.0  & 0.35 & 25 & 15 55 & 0.08 \\
    151307.75+605956.9 & 727    & 2.5 & 1.0 3.0 & 0.22 & 1  & 1 1  & 0.97 & 80.0  & 40.0 150.0  & 0.25 & 15 & 15 20 & 0.74 \\
    160004.33+550429.9 & 21285  & 0.0 & 0.0 2.5 & 0.02 & 4  & 3 14 & 0.04 & 100.0 & 30.0 150.0  & 0.15 & 20 & 15 60 & 0.08 \\
    160950.72+532909.5 & 372    & 3.0 & 1.0 3.0 & 0.08 & 4  & 3 8  & 0.30 & 80.0  & 10.0 80.0   & 0.21 & 15 & 15 20 & 0.76 \\
    161007.11+535814.0 & 27052  & 0.0 & 0.0 2.0 & 0.24 & 1  & 1 13 & 0.07 & 150.0 & 30.0 150.0  & 0.13 & 20 & 20 70 & 0.02 \\
    161238.27+532255.0 & 122721 & 0.0 & 0.0 2.0 & 0.07 & 1  & 1 14 & 0.02 & 150.0 & 30.0 150.0  & 0.11 & 15 & 15 60 & 0.06 \\
    163021.65+411147.1 & 17398  & 0.0 & 0.0 2.0 & 0.14 & 1  & 1 14 & 0.01 & 150.0 & 30.0 150.0  & 0.13 & 20 & 15 65 & 0.03 \\
    163425.11+404152.4 & 2607   & 3.0 & 0.5 3.0 & 0.10 & 2  & 1 10 & 0.14 & 80.0  & 30.0 150.0  & 0.18 & 15 & 15 25 & 0.52 \\
    163952.85+410344.8 & 11272  & 0.0 & 0.0 3.0 & 0.01 & 1  & 1 12 & 0.11 & 100.0 & 40.0 150.0  & 0.20 & 15 & 15 40 & 0.34 \\
    164016.08+412101.2 & 34911  & 0.0 & 0.0 2.5 & 0.02 & 1  & 1 14 & 0.00 & 100.0 & 30.0 150.0  & 0.15 & 15 & 15 60 & 0.09 \\
    172522.06+595251.0 & 1054   & 3.0 & 0.5 3.0 & 0.08 & 5  & 3 12 & 0.14 & 60.0  & 20.0 80.0   & 0.25 & 15 & 15 20 & 0.81 \\
    \enddata
    \tablecomments{For descriptions of \clumpy torus parameters, please see
      notes to Table~\ref{tab:torus_best_fit_without_bb}. Larger $D_{KL}$
      value implies that the parameter distribution is more peaked, and the
      respective parameter is better constrained. Smaller number of accepted
      models also imply better fits. These statistics are generated by
      selecting models that differ from the best-fit model SED by 10\%,
      relaxing this criterion flattens the parameter distributions.}
    \label{tab:torus_par_stats}
  \end{deluxetable}
  
  \begin{deluxetable}{lr|rrr|rrr|rrr}
    \setcounter{table}{6}
    \tabletypesize{\tiny}
    \tablewidth{0pt}
    \tablecolumns{11}
    \tablecaption{Statistics on the parameter distributions  -- Continued.}
    \tablehead{
      \colhead{SDSS ID} &
      \colhead{Number} &
      \multicolumn{3}{c}{$Y$} &
      \multicolumn{3}{c}{$i$} &
      \multicolumn{3}{c}{$T_{BB}({\rm K})$} \\
      \colhead{} &
      \colhead{of models} &
      \colhead{Mode} &
      \colhead{90\%} &
      \colhead{$D_{KL}$} &
      \colhead{Mode} &
      \colhead{90\%} &
      \colhead{$D_{KL}$} &
      \colhead{Mode} &
      \colhead{90\%} &
      \colhead{$D_{KL}$}
    }
    \startdata
    095047.47+480047.3 & 668    & 70       & 40 100 & 0.09 & 90 & 70 90 & 0.55 & 800 900   & 800 1300  & 0.57 \\
    100401.27+423123.0 & 883    & 5        & 5 90   & 0.02 & 70 & 30 90 & 0.20 & 900 1000  & 900 1500  & 0.35 \\
    103931.14+581709.4 & 1857   & 100      & 10 100 & 0.01 & 90 & 60 90 & 0.34 & 1100 1200 & 1000 1200 & 0.77 \\
    104114.48+575023.9 & 7911   & 100      & 10 100 & 0.00 & 70 & 30 90 & 0.16 & 1200 1300 & 800 1300  & 0.52 \\
    104155.16+571603.0 & 1392   & 100      & 10 100 & 0.01 & 80 & 70 90 & 0.48 & 1100 1200 & 1000 1300 & 0.65 \\
    104355.49+562757.1 & 3888   & 100      & 20 100 & 0.02 & 90 & 60 90 & 0.37 & 800 900   & 800 1000  & 0.69 \\
    105001.04+591111.9 & 27260  & 5        & 5 90   & 0.00 & 70 & 40 90 & 0.17 & 1300 1400 & 1200 1400 & 0.78 \\
    105153.77+565005.7 & 24033  & 100      & 10 100 & 0.00 & 70 & 30 90 & 0.16 & 1200 1300 & 900 1400  & 0.50 \\
    105447.28+581909.5 & 63     & 40 50    & 20 100 & 0.07 & 60 & 50 90 & 0.31 & 1300 1400 & 1200 1500 & 0.70 \\
    105951.05+090905.7 & 28     & 5        & 5 80   & 0.12 & 90 & 90 90 & 1.00 & 900 1000  & 800 1100  & 0.66 \\
    132120.48+574259.4 & 6113   & 60       & 40 100 & 0.11 & 80 & 70 90 & 0.45 & 900 1000  & 900 1200  & 0.59 \\
    142730.19+324106.4 & 281    & 70       & 20 100 & 0.04 & 90 & 70 90 & 0.59 & 800 900   & 800 1100  & 0.59 \\
    142954.70+330134.7 & 3537   & 70       & 10 100 & 0.00 & 70 & 40 90 & 0.24 & 900 1000  & 800 1400  & 0.35 \\
    143102.94+323927.8 & 21     & 40 80 90 & 10 90  & 0.06 & 80 & 80 90 & 0.79 & 1000 1100 & 900 1300  & 0.52 \\
    143605.07+334242.6 & 3765   & 100      & 20 100 & 0.02 & 70 & 60 90 & 0.29 & 1100 1200 & 900 1400  & 0.41 \\
    151307.75+605956.9 & 727    & 5        & 5 90   & 0.02 & 90 & 20 90 & 0.12 & 1100 1200 & 1000 1700 & 0.36 \\
    160004.33+550429.9 & 21285  & 70       & 10 100 & 0.00 & 80 & 40 90 & 0.21 & 1100 1200 & 800 1200  & 0.51 \\
    160950.72+532909.5 & 372    & 5        & 5 90   & 0.01 & 80 & 80 90 & 0.72 & 1300 1400 & 1200 1400 & 0.72 \\
    161007.11+535814.0 & 27052  & 90       & 10 100 & 0.02 & 50 & 10 90 & 0.02 & 1100 1200 & 1000 1200 & 0.71 \\
    161238.27+532255.0 & 122721 & 100      & 20 100 & 0.02 & 90 & 40 90 & 0.19 & 1100 1200 & 1100 1300 & 0.68 \\
    163021.65+411147.1 & 17398  & 100      & 20 100 & 0.03 & 90 & 30 90 & 0.15 & 1100 1200 & 1100 1300 & 0.74 \\
    163425.11+404152.4 & 2607   & 5        & 5 90   & 0.00 & 80 & 70 90 & 0.42 & 1200 1300 & 1000 1400 & 0.52 \\
    163952.85+410344.8 & 11272  & 80       & 10 100 & 0.00 & 80 & 30 90 & 0.18 & 1300 1400 & 1100 1400 & 0.71 \\
    164016.08+412101.2 & 34911  & 20       & 10 100 & 0.01 & 90 & 40 90 & 0.22 & 1200 1300 & 1100 1600 & 0.42 \\
    172522.06+595251.0 & 1054   & 5        & 5 90   & 0.01 & 90 & 80 90 & 0.65 & 1200 1300 & 1200 1400 & 0.73 \\
    \enddata
    \tablecomments{Description of parameters: $Y$: the ratio of outer to inner
      (sublimation) radius of the torus.; $i$: inclination of line-of-sight of
      the observer; $T_{BB}$: Temperature of the blackbody component in
      Kelvin. Larger $D_{KL}$ value implies that the parameter distribution is
      more peaked, and the corresponding parameter is better constrained.
      Smaller number of accepted models also imply better fits.}
  \end{deluxetable}
  
\end{document}